\documentclass[useAMS,usenatbib]{mn2e}
\usepackage{natbib}
\usepackage{graphicx}
\usepackage{savesym}
\usepackage{listings}
\usepackage[intlimits]{amsmath}%option changes placement of integral limits
\savesymbol{iint}
\usepackage{txfonts}
\usepackage{bm}
\restoresymbol{TXF}{iint}
\usepackage{amssymb}
\usepackage{amsmath}
\usepackage{tabularx} 
\usepackage{color}
\usepackage{soul}
\usepackage{comment}

\def\ba{\begin{eqnarray}}
\def\bam{\begin{array}}

\def\be{\begin{equation}}

\def\B {\overline}

\def\Br{\B r}

\def\de{\delta}

\def\ea{\end{eqnarray}} 
\def\ee{\end{equation}}

\def\fr{\frac}

\def\ha{\frac{1}{2}~}

\def\ts{\textstyle}

\def\1{{\it one}}

\def\2{{\ts{\ha}\!}}

\def\3 {\ts{\frac{1}{3}\!}}
\def\4{\ts{\fr{1}{4}\!}}

\def\be{\begin{equation}}
\def\ee{\end{equation}}
\def\ba{\begin{eqnarray}}
\def\ea{\end{eqnarray}}
\def\go{\mathrel{\raise.3ex\hbox{$>$}\mkern-14mu
             \lower0.6ex\hbox{$\sim$}}}
\def\lo{\mathrel{\raise.3ex\hbox{$<$}\mkern-14mu
             \lower0.6ex\hbox{$\sim$}}}

\begin{document}

\title[Coupled axisymmetric pulsar magnetospheres]{Coupled axisymmetric pulsar magnetospheres}

\author[K.N. Gourgouliatos \& D. Lynden-Bell]{{Konstantinos N. Gourgouliatos$^{1}$ \& Donald Lynden-Bell\thanks{Deceased}}
\vspace{0.4cm}\\
\parbox{\textwidth}{$^{1}$Department of Mathematical Sciences, Durham University, Durham,  DH1 3LE, UK, Konstantinos.Gourgouliatos@durham.ac.uk}}

\maketitle

\label{firstpage}
\pagerange{\pageref{firstpage}--\pageref{lastpage}}

\begin{abstract}
We present solutions of force-free pulsar magnetospheres coupled with a uniform external magnetic field aligned with the dipole magnetic moment of the pulsar. 
The inclusion of the uniform magnetic field has the following consequences: The equatorial current sheet is truncated to a finite length, the fraction of field lines that are open increases, and the open field lines are confined within a cylindrical surface instead of becoming radial.  A strong external magnetic field antiparallel to the dipole allows for solutions where the pulsar magnetic field is fully enclosed within an ellipsoidal surface. Configurations of fully enclosed or confined magnetospheres may appear in a double neutron star (DNS) where one of the components is a slowly rotating, strongly magnetised pulsar and the other a weakly magnetised millisecond pulsar. Depending on the geometry, twisted field lines from the millisecond pulsar could form an asymmetric wind. More dramatic consequences could appear closer to merger: there, the neutron star with the weaker magnetic field may undergo a stage where it alternates between an open and a fully enclosed magnetosphere releasing up $10^{38}$erg. Under favourable combinations of magnetic fields DNSs can spend up to 10$^3$ yr in the coupled phase, implying a Galactic population of $0.02$ systems. Potential observational signatures of this interaction are discussed including the possibility of powering recurring Fast Radio Bursts (FRBs). We also note that the magnetic interaction cannot have any significant effect on the spin evolution of the two pulsars. 
\end{abstract}

\date{\today}


\section{Introduction}

The recent discovery of gravitational waves from coalescing neutron stars in the GW170817 event \citep{Abbott:2017a} has permitted a clear view of this extreme phenomenon across the electromagnetic spectrum \citep{Abbott:2017b}. The electromagnetic counterpart observed simultaneously and after the gravitational wave detection has confirmed the predictions of the kilonova model and have set constraints on the formation of a relativistic jet \citep{Metzger:2010, Metzger:2012, Cowperthwaite:2017, Nicholl:2017, Margutti:2017, Alexander:2017}. Even before the merger, though, the magnetospheres of neutron stars will be interacting and modify each other's structure giving rise to precursor events \citep{Palenzuela:2013, Piro:2012, Tsang:2012, Tsang:2013} possibly associated with radiation received before the prompt emission of a short gamma-ray burst \citep{Troja:2010}. The magnetic interaction of the two neutron stars has been simulated for small separations just before the merger takes place \citep{Rezzolla:2011, Etienne:2012, Paschalidis:2013, Ponce:2014}. While the interaction of the magnetospheres shortly before and during the merger will definitely be the most dramatic, a window of opportunity of observable effects may occur at earlier times when the two magnetospheres are coupled. 

DNSs can form through the recycled pulsar channel \citep{Tauris:2017, Bhattacharya:1991}. In this scenario the older member of the binary is a millisecond pulsar with a relatively weak magnetic field ($\sim 10^{9}$~G), whereas the younger one has a magnetic field a few orders of magnitude higher and spin slower. Due to gravitational wave decay their orbit shrinks. It is then possible that the strength of the magnetic field of the younger pulsar at the neighbourhood of the millisecond pulsar to be higher than that of the millisecond pulsar itself. In such a configuration, the magnetosphere of the millisecond pulsar will be drastically modified. 

As of now 16 DNS systems and candidates have been identified \citep{Tauris:2017, Stovall:2018}. Among them, the double pulsar J0737-3039A/B is expected to coalesce in 86 Myr \citep{Stairs:2004}. As the two neutron stars move closer they will reach a point where the two magnetospheres interact. Even then, the characteristic length-scales of the system (light-cylinders and orbital separation) will still differ by a few orders of magnitude. Because of this, direct numerical simulation may not be the optimal approach. Instead, one could consider the effect of a uniform magnetic field onto a pulsar magnetosphere. In this approach the uniform external field represents the field of the young pulsar that modifies the millisecond pulsar magnetosphere. 

The \cite{Goldreich:1969a} pulsar model describes the magnetic field configuration of an aligned rotating magnetic dipole. This solution corresponds to a force-free magnetosphere that transitions to a wind at great distances. The magnetosphere contains an equatorial current sheet extending from the light-cylinder radius ($R_{LC}$) to infinity, where  poloidal and toroidal electric currents flow. Numerical solutions of the axisymmetric problem postulate the presence of the current sheet on the equatorial plane through the imposed boundary conditions \citep{Contopoulos:1999} and then solve for half the volume, either north or south of the equator, taking advantage of the symmetry of the problem. Solutions of the pulsar equation are still feasible if the current sheet continues within the light-cylinder. These solutions hold when the open magnetic field lines rotate differentially instead of corotating with the neutron star \citep{Contopoulos:2005, Timokhin:2006}. The current sheet in these solutions extends to infinity as well.

Solving  the pulsar equation for the entire volume, without imposing any condition on the equatorial plane, leads to a strong, but finite, equatorial current. Indeed, due to the inevitable numerical dissipation of the arithmetic scheme, the formation of infinitesimal current sheets is not feasible \citep{Komissarov:2006}. Oblique rotators are even more complicated, not permitting the system to relax to a stationary state. In these systems the field is solved for under the approximation of force-free electrodynamics. Such non-axisymmetric configurations still contain thin currents layers emanating from the light-cylinder and extending to large distance  from the pulsar \citep{Spitkovsky:2006, Kalapotharakos:2009, Kalapotharakos:2012}. Beyond the light-cylinder the system is dominated by Poynting flux transitioning eventually to a low magnetisation hydrodynamical wind \citep{Vlahakis:2004}.

The influence of a strong external magnetic field may modify drastically the pulsar magnetosphere and wind. DNSs will affect each other either via winds at large separations  or through direct interaction of their magnetospheres once they are close to each other.  
In a previous paper  \citep{Gourgouliatos:2011a} 
we have provided a heuristic  description of the magnetic field of a rotating dipole coupled to a uniform external magnetic field. Here, we revisit this problem through exact analytical and numerical solutions. As a key element of the pulsar magnetosphere is its infinite current sheet, we start our discussion from the modification of a current sheet arising by a split monopole, once an external uniform magnetic field is included. This configuration is then used as a guide for the numerical solution of a system where a relativistically rotating dipole magnetic field is embedded inside a uniform magnetic field. 

The plan of the paper is as follows. In section 2 we solve analytically and numerically a model problem of a split monopole embedded in a uniform external field. In section 3, we solve numerically the pulsar equation embedded in a uniform magnetic field. We present the implications of these solutions to magnetically interacting DNSs in section 4. We discuss our results in section 5. We conclude in section 6.

\section{A Model Problem}

Embedding the pulsar magnetosphere into a uniform background field should truncate the infinite current sheet to a finite length. As this is going to be an assumption that will be included in the numerical calculation through the boundary conditions, we need to check its validity. To do so, we solve first the problem of the force-free equilibrium of a split monopole in a uniform field. We cannot just add the fields as that gives a net field in the current sheet which destroys the force-free nature of the field. At large distances the externally imposed uniform field must dominate while at small distances the split monopole does. The field lines connected to the source and those of the uniform field are divided by a separatrix.

\subsection{Sphere superposed with uniform field}

	Consider a sphere of radius $a$ that is penetrated along its axis by a thin magnet of length $2a$ so that the north pole is of strength $M$ and the south pole is of strength $-M$. The sphere away from its axis is perfectly conducting and allows no field to penetrate it. This sphere is placed in an external field whose strength is just such that it annihilates the field on the sphere's equator. To solve this problem we superpose a field $B_0$ that is straight at infinity but can not penetrate the sphere, with the field of the magnet through the sphere that can not penetrate it away from the poles. We use spherical coordinates $(r, \theta, \phi)$.
	The first field is given by $\nabla\chi_1$ where
\be
 \chi_1= -B_0 \left[\frac{r}{a}+\frac{1}{2} \left(\frac{a}{r}\right)^2\right]\cos \theta,
\ee
which is by construction harmonic and does not penetrate the sphere since $\partial\chi/\partial r$
vanishes at $r=a$.
	The second field is $\chi_2=\sum\limits_{n=0}^{\infty} a_{2n+1}(a/r)^{2n+2}P_{2n+1}(\mu) $, with $\mu=
	\cos\theta$, where the $\delta$-functions
$2 M/a^2 \de(|\mu|-1)$ at each pole and the zero radial gradient elsewhere on the sphere give,
\begin{eqnarray}
4 M/a^2 &=&[2(2n+2)/(4n+3)]a_{2n+1}/a\nonumber \\
a_{2n+1}&=& [(4n+3)/(n+1)]M/a
\end{eqnarray}
With $\chi=\chi_1+\chi_2$ and the condition that $\partial \chi/\partial \theta= 0 $ on the equator
we find $P_1'(0)=1$ and
\begin{eqnarray}
(3/2)B_0=(M/a)\sum\frac{4n+3}{n+1}P_{2n+1}'(0)\label{B_0}
\end{eqnarray}
where $P_{2n+1}'(0)=(-1)^n[(n+\2)(n-\2)...3/2]/n!$. 

The field of the model problem is now fully determined by $\chi(r,\theta)$. \\

The sum involving the Legendre Polynomials can be evaluated in terms of an analytical function. The sum giving the second term of the potential is
\be 
\chi_{2}=\frac{4 M}{a}\sum\limits_{n=0}^{\infty} \frac{4n+3}{4n+4}\left(\frac{a}{r}\right)^{2n+2}P_{2n+1}(\mu)\,,
\ee
setting $x=(a/r)$ we can write it in the form
\be
\chi_{2}=\frac{4M}{a}\sum\limits_{n=0}^{\infty}\left(1-\frac{1}{4n+4}\right)x^{2n+2}P_{2n+1}(\mu)\,.
\ee
The generating function for the Legendre polynomials is: 
\be
G(\mu,x)=(1-2\mu x +x^2)^{-1/2}=\nonumber  \\
\sum\limits_{n=0}^{\infty}x^{n}P_{n}(\mu)\,,
\ee
to keep only the odd terms in the sum we subtract $G(-\mu, x)$ and by multiplication with $2Mx/a$ we take the first term of the $\chi_{2}$:
\begin{eqnarray}
\frac{2M}{a}x[G(\mu, x)-G(-\mu, x)]=\frac{4M}{a}\sum\limits_{n=0}^{\infty}x^{2n+2}P_{2n+1}(\mu) =\nonumber  \\
\frac{2M}{a}x[(1-2\mu x+x^2)^{-1/2}-(1+2\mu x +x^2)^{-1/2}]\,.
\end{eqnarray}
To take the second term of the sum we integrate $G(\mu, x)-G(-\mu, x)$ in $x$ and we multiply by $M/(2a)$. The result is 
\begin{eqnarray}
\frac{4M}{a}\sum\limits_{n=0}^{\infty}\frac{x^{2n+2}}{4n+4}P_{2n+1}(\mu)=\nonumber  \\
\frac{M}{a}\int_{0}^{x}[(x-2\mu x+x^2)^{-1/2}-(x+2\mu x+x^2)^{-1/2}]dx = \nonumber\\ 
\ln\frac{1+\mu}{1-\mu} + \ln\frac{x-\mu +\sqrt{1-2\mu x +x^2}}{x+\mu +\sqrt{1+2\mu x +x^2}}\,.
\end{eqnarray}
Therefore the potential $\chi$ is
\begin{eqnarray}
\chi= -B_{0}(x^{-1}+\frac{1}{2}x^{2})\mu+ \nonumber \\
\frac{M}{\alpha}\Big\{ 2x[(1-2\mu x+x^{2})^{-1/2} -(1+2\mu x+x^{2})^{-1/2}]+\nonumber \\
\ln \Big[\Big(\frac{1-\mu}{1+\mu}\Big) \frac{x+\mu + (1+2\mu x +x^{2})^{1/2}}{x-\mu+(1-2\mu x + x^{2})^{1/2}}\Big]\Big\}
\label{POTENTIAL}
\end{eqnarray}
Then for $\mu=0$, $x=1$ and by virtue of Equation (\ref{B_0}) we find that $B_{0}= 0.552 M/a$. 

\begin{figure}
\includegraphics[width=0.5\columnwidth]{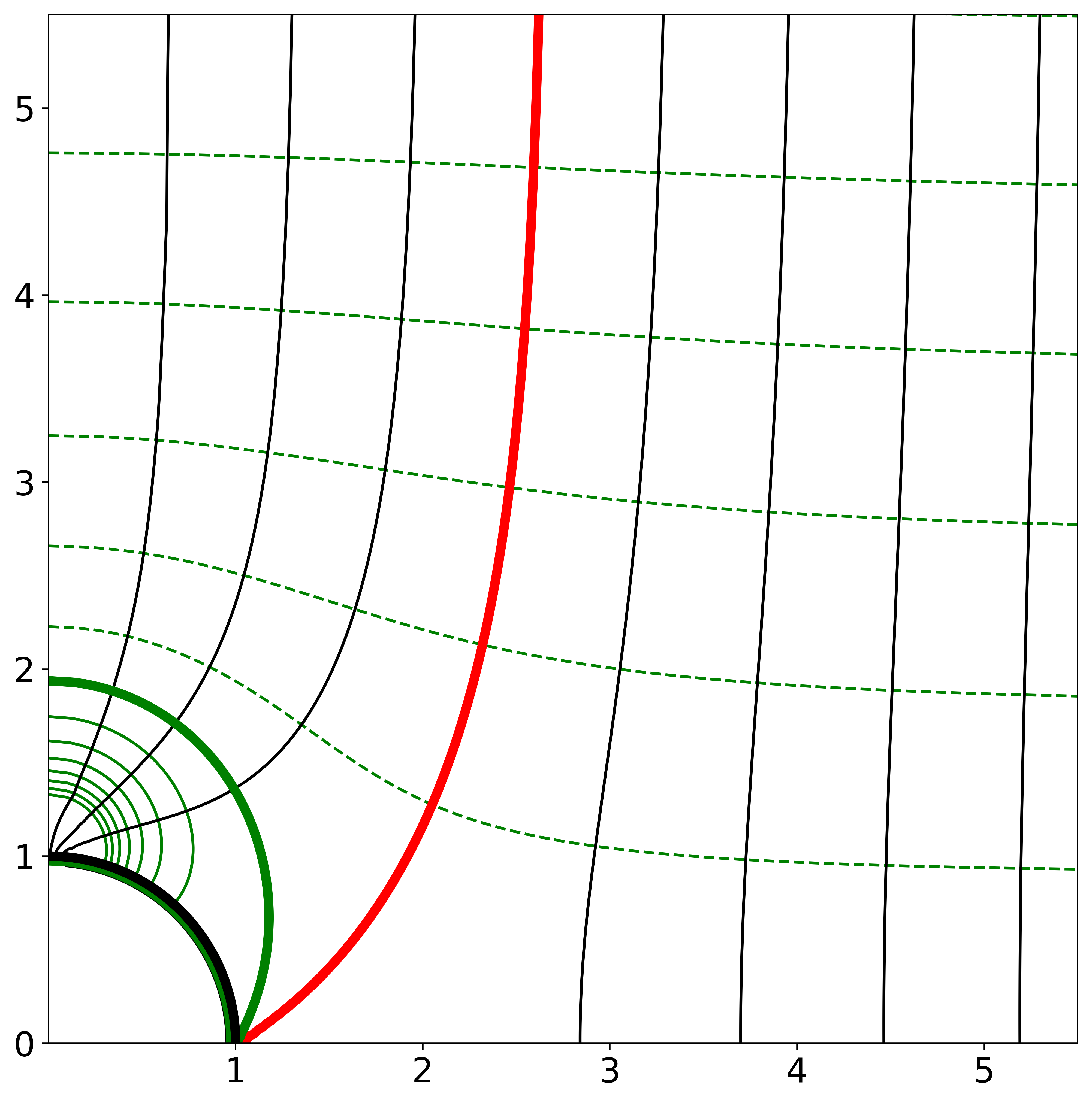}
\includegraphics[width=0.5\columnwidth]{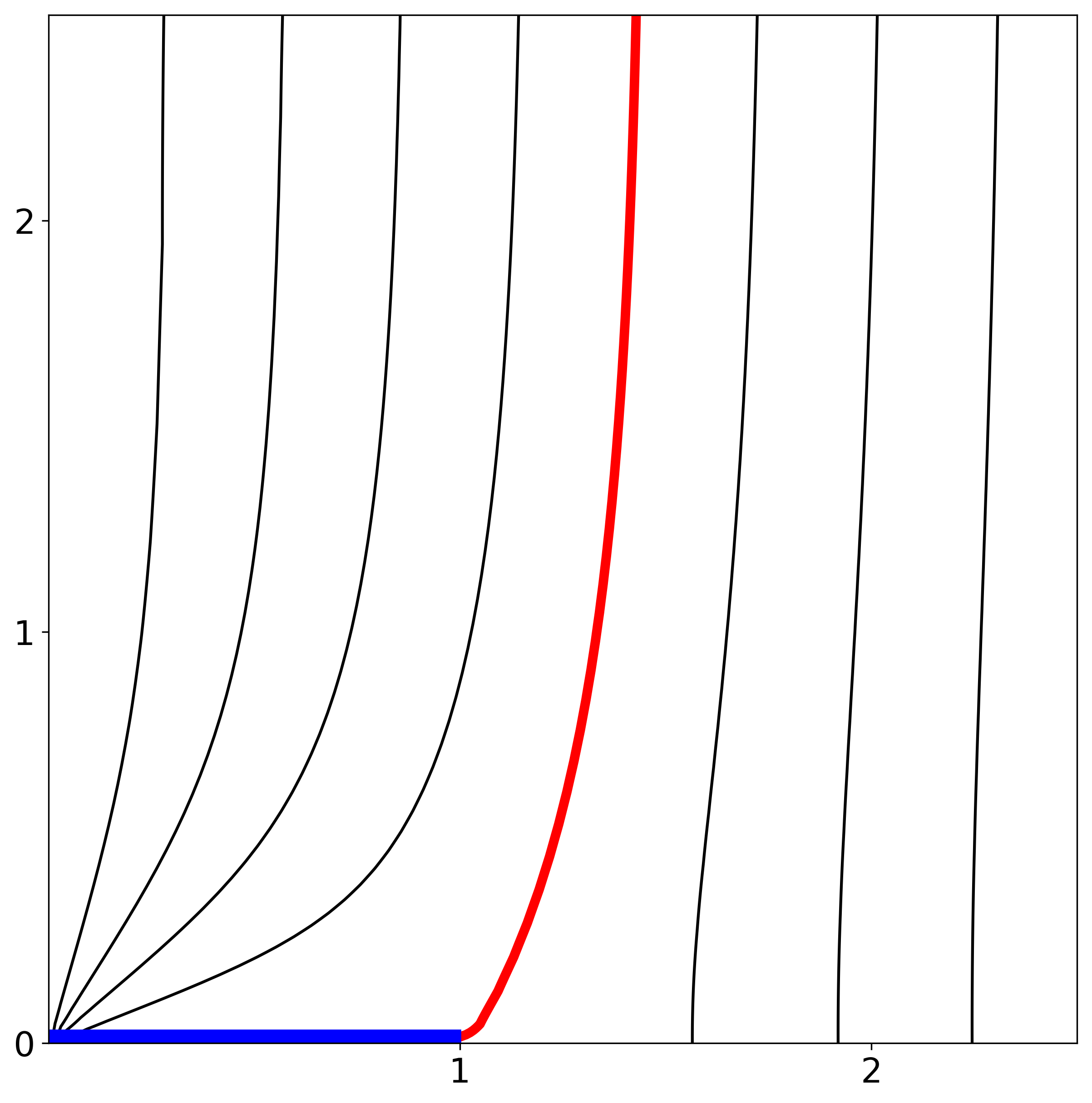}
\caption{Left Panel. Equipotential curves ($\chi=$const.) given by equation \ref{POTENTIAL} plotted in green outside of a sphere of radius $a=1$. The solid green lines have $\chi(r,\mu)>\chi(1, 0)$, the dashed  $\chi(r,\mu)<\chi(1, 0)$ and the thick green line corresponds to $\chi(r,\mu)=\chi(1, 0)$. The black lines are the field lines ($\Psi=$const.) found numerically. The potential lines are normal to the field lines as required. The separatrix diving the field lines emanating from the pole to the external ones is plotted in red. Right Panel. The field lines (black solid lines) arising from the superposition of the split monopole and the external field. The disc where the current sheet is located in shown in blue and the separatrix in red.}
\label{Figure:1}
\end{figure}

We have also solved this problem numerically. We express the magnetic field in terms of the poloidal flux $\bm{B}=\nabla \Psi \times \nabla \phi$, in spherical coordinates. The boundary conditions are $\Psi(r, 0)=0$, $\Psi(a, \theta)=1$ for $\theta>0$, $\frac{\partial \Psi}{\partial \theta}|_{\theta=\pi/2}=0$ and $\frac{\partial \Psi}{\partial r}|_{r=r_{b}}=r_{b}\sin^{2}\theta B_{0}$, where $r_{b}$ is the radius of the numerical box. The value of $B_{0}$ is chosen so that the separatrix passes from $(a, 0)$, the equator of the sphere. If a smaller value for $B_{0}$ were chosen there would be field lines emanating from the poles of the sphere crossing the equator and the separatrix crosses the equatorial plane at some $r>a$. Alternatively if a larger value for $B_{0}$ were chosen there would be field lines from the external field connecting onto the sphere. Using a $200\times 200$ grid in $r$ and $\mu=\cos\theta$ with $r_{b}/a=10$ we find that the numerical value of $B_{0}$ deviates $< 1\%$ from the analytical.The results of the numerical and the analytical solutions appear in the left panel of Figure \ref{Figure:1}. 

\subsection{Split monopole in a uniform field}
	We now return to the corresponding problem in spheroidal coordinates  $\Br,\B\theta$. The roots for
 $\tau$ of the quadratic that follows are $\Br^2$ and $-a^2\cos^2\B\theta$ where $R,z$ are cylindrical polar coordinates.
\be
\frac{R^2}{\tau+a^2}+\frac{z^2}{\tau}=1;~~=>~~\tau^2-(R^2+z^2-a^2)\tau-a^2z^2=0
\ee
so $\Br=0$ is the disc of radius $a$ and greater values give confocal oblate spheroids.  $\B\theta=const$
are confocal hyperboloids of one sheet.
\be
z=\Br\cos\B\theta;~~~~~~~R^2=(\Br^2+a^2)\sin^2\B\theta
\ee
In these coordinates the metric is
\be
ds^2=\left(\Br^2+a^2\cos^2\B\theta\right)\left[\frac{d\Br^2}{\Br^2+a^2}+d\B\theta^2\right]+\left(\Br^2+a^2\right)\sin^2\B\theta d\phi^2.
\ee
$\nabla^2\chi=0$ separates in these coordinates yielding solutions that die at large $\Br$
\be
\chi=\sum a_lQ_l(\Br/ia)P_ l(\B\mu)
\ee
where $\B\mu=\cos\B\theta$ and both $P_l(\B\mu)$ and $Q_l(u)$ satisfy Legendre's equation although  $u$ is pure imaginary with $iu>0$. As before the problem only needs the odd $l$ solutions and the dipole solution is
\begin{eqnarray}
\chi_1= -B_0[\Br-Q_1(\Br/ia)/Q_{1,0}']\cos\B\theta ;\nonumber \\
Q_1(\Br/ia)=(\Br/a) \tan^{-1}(a/\Br)-1;~Q_{1,0}'=\2\pi/a.
\end{eqnarray}
The requirement that the ``radial" gradient must be given by the delta functions at the poles gives

\be
4M/a^2=[2/(4l+3)]a_{2l+1}Q'_{2l+1}(0)/(ia),
\label{MAB}
\ee
 
 At zero the $Q_{2l+1}'$ and the $Q_{l}$ obey the recurrence relations $Q_{l}'=lQ_{l-1}$ and 
 $lQ_{l}=-(l-1)Q_{l-2}$ also $Q_0(\Br/ia)=i\tan^{-1}(a/\Br)->i\pi/2$ so these suffice to
 determine the $a_{2l+1}$. 
 Finally we must determine $a$ from the requirement that there be a neutral point of the whole field at
 $\B\mu=0,\Br=0$. This gives
  \be
  2aB_0/\pi=\sum a_{2l+1}Q_{2l+1}(0)P_{2l+1}'(0)
  \ee
  and by equation \ref{MAB} we find that 
  \begin{eqnarray}
  a^2B_{0}/M={\rm const.}
  \label{SPLIT_MON}
  \end{eqnarray}
   Thus the square of the radius of the current sheet is inversely proportional to the magnetic field. Note the difference between the sheet and the sphere utilised in the model problem: in the latter for a given $M$ the radius of the sphere was inversely proportional to the strength of the external uniform field. 
   
  Again at $0$ the $Q_{2l+1}$ can be determined from the recurrence relations and the expression for
  $Q_1$  given above. Unlike the sphere problem we have not obtained the solution in closed form. Thus the complete solution for $\chi(\Br,\B\mu)$ is determined as a sum. We plot the magnetic field in Figure \ref{Figure:1} right panel, found numerically using the method  described in the model problem.

The model problem discussed above demonstrates than an infinite current sheet will be truncated and become finite once a uniform field is superposed.

\section{Axisymmetric pulsar magnetosphere in a uniform magnetic field}
\label{NUM_SOL}

\begin{figure*}
\includegraphics[width=0.68\columnwidth]{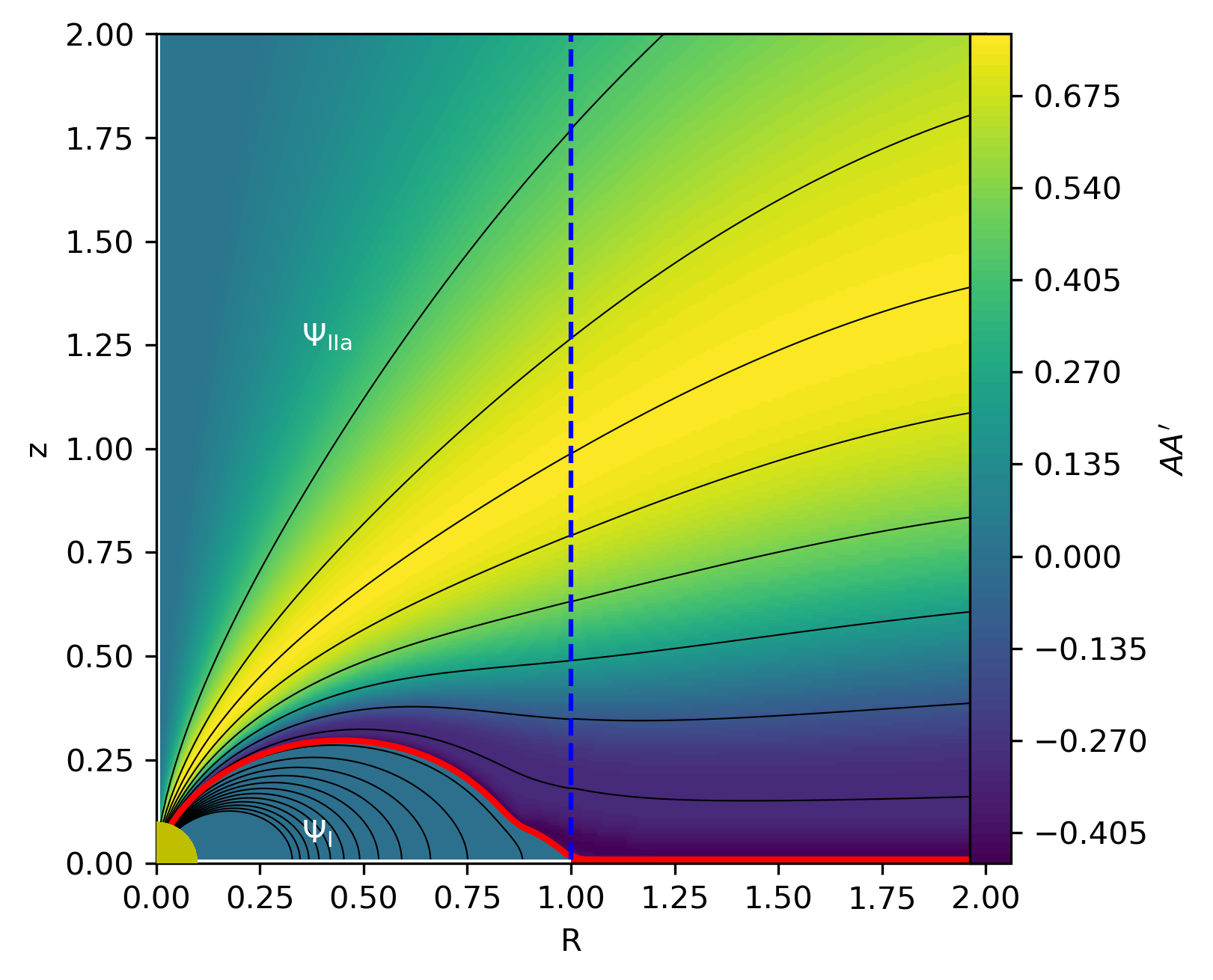}
\includegraphics[width=0.68\columnwidth]{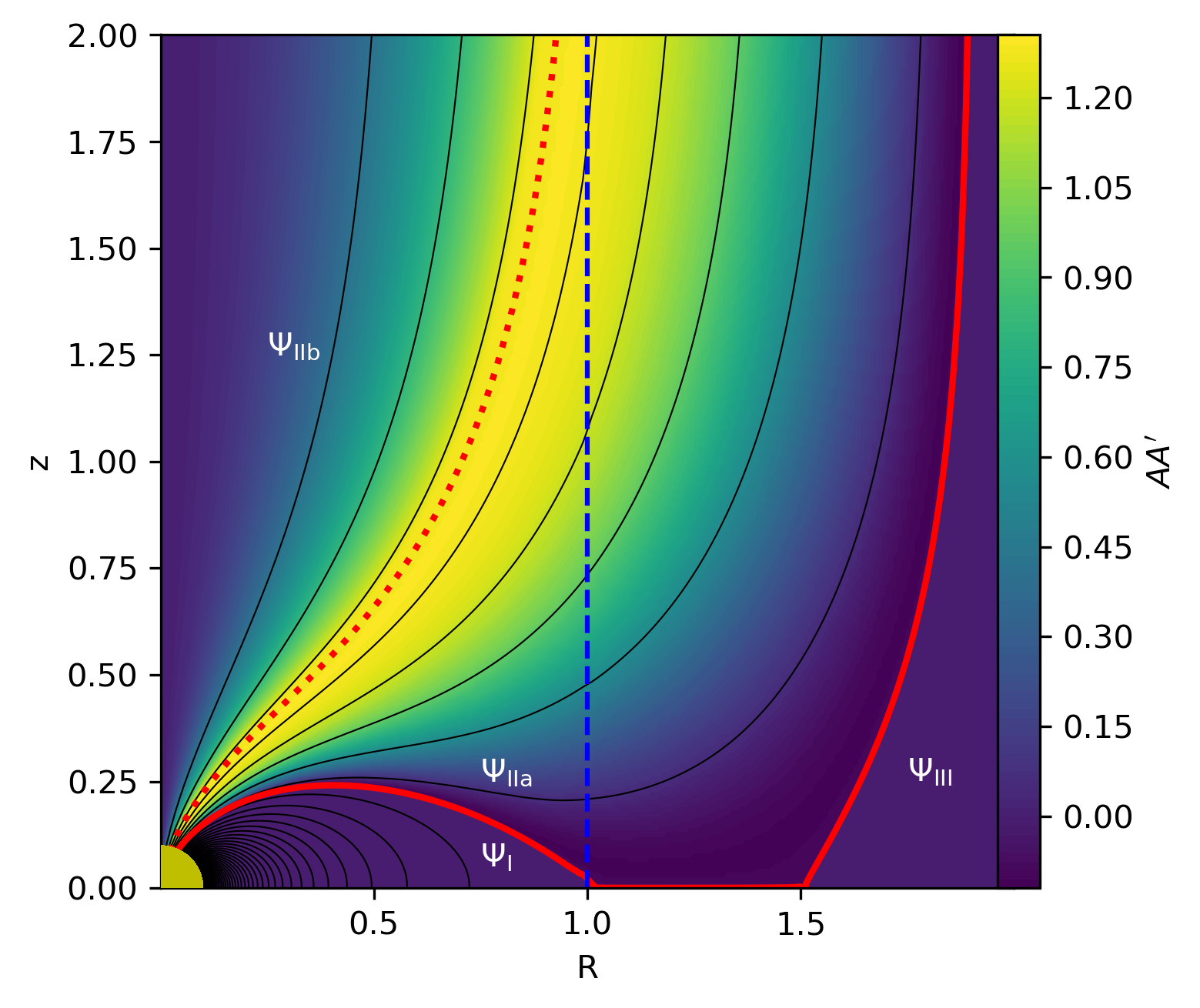}
\includegraphics[width=0.68\columnwidth]{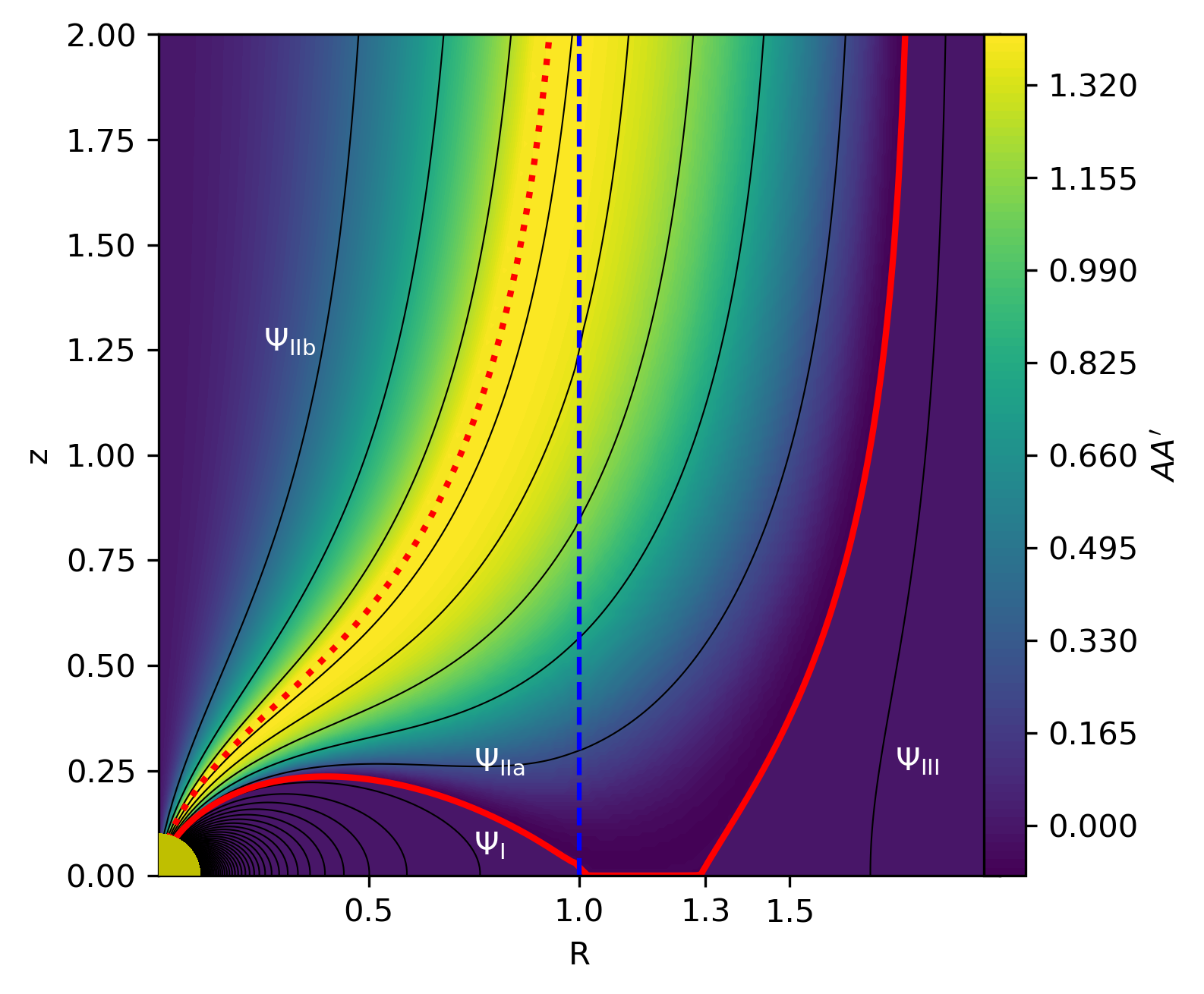}
\includegraphics[width=0.68\columnwidth]{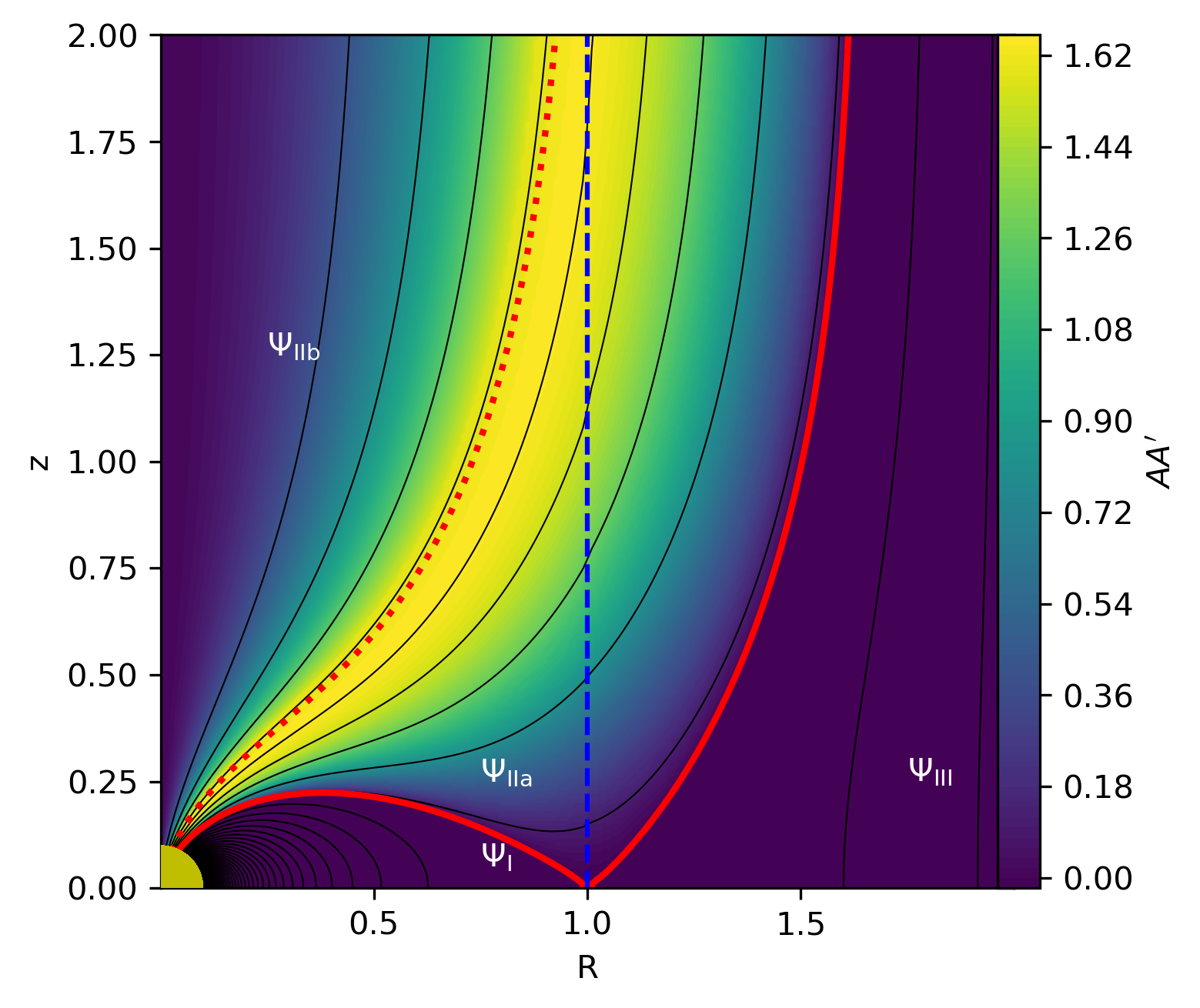}
\includegraphics[width=0.68\columnwidth]{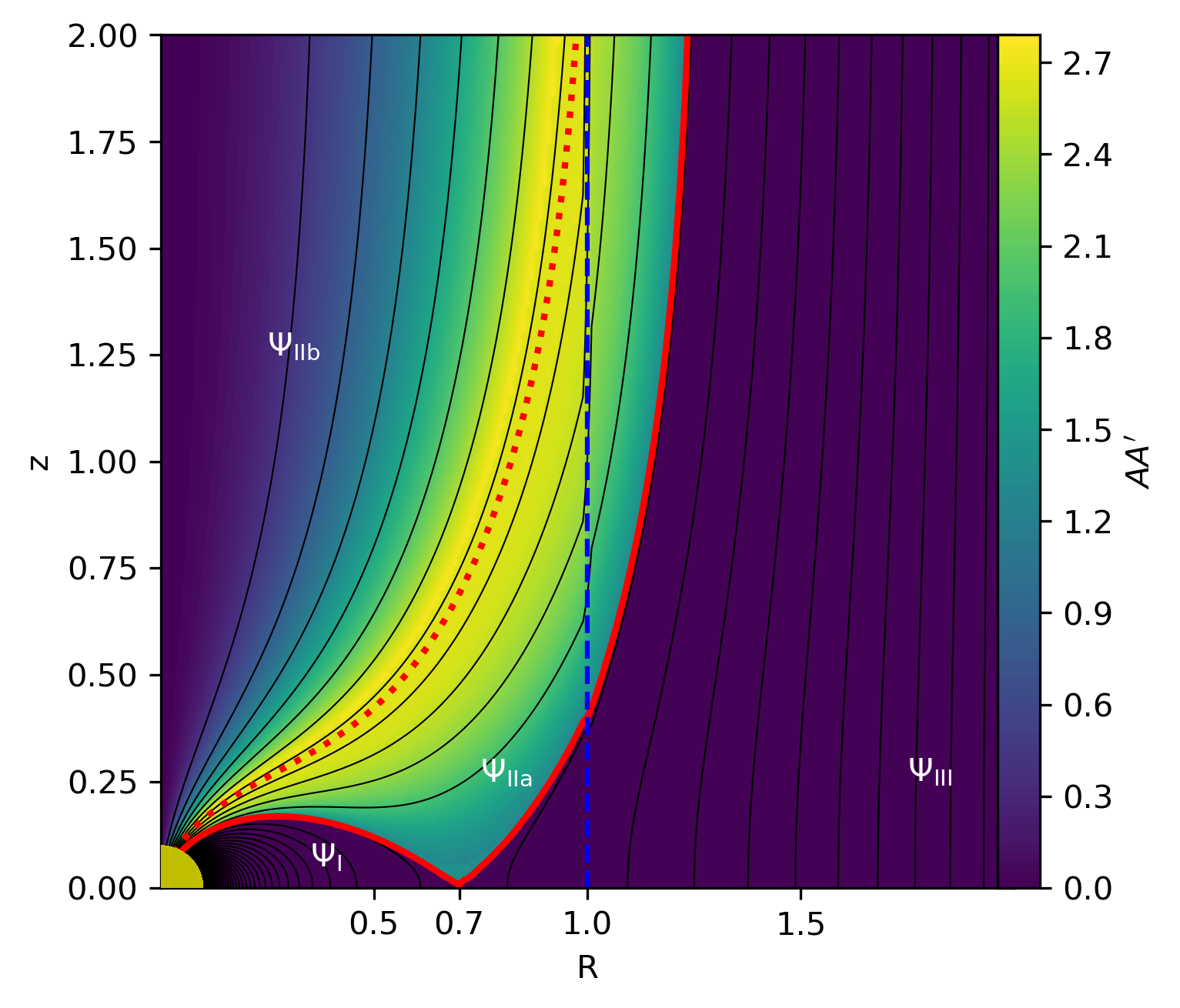}
\includegraphics[width=0.68\columnwidth]{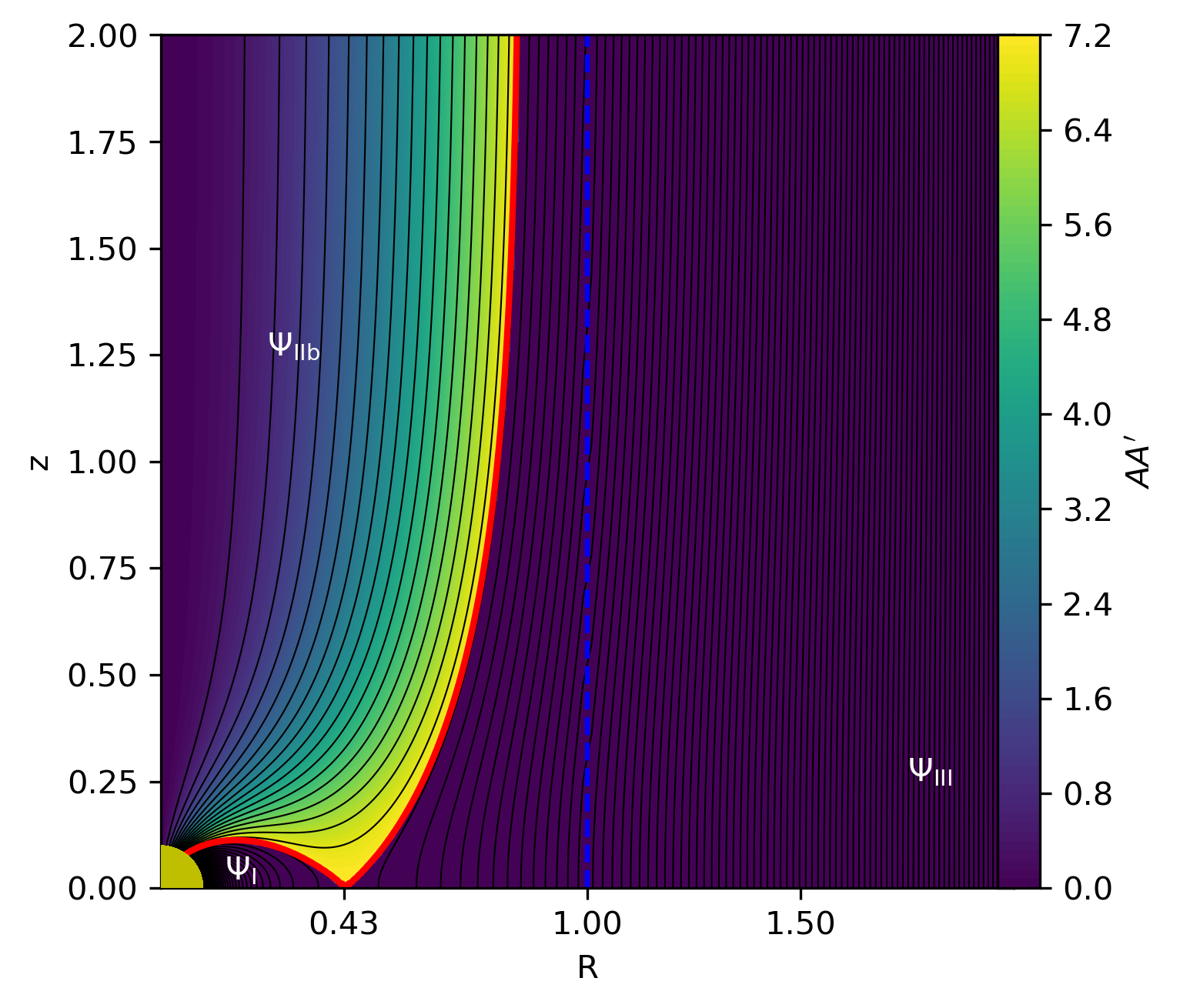}
\caption{Magnetic field structure for the isolated force-free magnetosphere P0 and runs A1 and A2 (top row from left to right) and runs A3, A4 and A6 bottom left to right. Black lines correspond to surfaces of constant $\Psi$, the colour scale is $AA^{\prime}$. The thick red line is the last open field line and the dashed blue line is the light-cylinder. The dotted red line is the boundary between regions IIa and IIb. Notice the absence of regions IIb and III in the isolated pulsar solution (P0) and  the absence of region IIa in the A6 solution, where no field lines starting from the pulsar cross the light cylinder.}
\label{Figure:2}
\end{figure*}

Consider an axially symmetric magnetic field in cylindrical polar coordinates $(R,\, \phi,\, z)$
\begin{eqnarray}
\bm{B}=\nabla \Psi \times \nabla \phi + A\nabla \phi\,,
\end{eqnarray}
where $\Psi=\Psi(R,z)$ and $A=A(R,z)$. The force-free equilibrium of this system satisfies the partial differential equation:
\begin{eqnarray}
\left(1-R^2\right)\left(\frac{\partial^2 \Psi}{\partial R^2}-\frac{1}{R}\frac{\partial \Psi}{\partial R}+\frac{\partial^2 \Psi}{\partial z^2}\right)-2R\frac{\partial \Psi}{\partial R} =- AA^{\prime}\,,~~
A=A(\Psi)\,,
\end{eqnarray} 
where we set the light-cylinder radius to be at $R_{LC}=c/\Omega=1$ and we use dimensionless units $c=1$, with $c$ the speed of light and $\Omega$ the angular frequency of the pulsar. The neutron star lies at the origin of the coordinates. Its magnetic moment and axis of rotation are along the $z$ direction. The magnetic field of an isolated pulsar satisfies the following boundary conditions: The field is radial at infinity 
\be
\lim_{R^2+z^2\to \infty}\left(B_{R}/B_{z}\right)=\arctan\theta
\ee
where $\theta$ is the polar angle measured from the $z$ axis. The closed field lines cross the equatorial plane perpendicularly; an equatorial current sheet forms outside the light-cylinder at $z=0$; and there is no radial field at $R=0$. These are summarised below:
\be
B_R(R<1, 0)=0\,, ~B_{z}(R>1, 0)=0\,,~B_{R}(0,z)=0\,.
\ee
We set the neutron star radius $r_{NS}=0.1$. $\Psi$ on the star  surface is
\begin{eqnarray}
\Psi=\frac{MR^2}{\left(R^2+z^2\right)^{3/2}}\,, ~R^2+z^2=r_{NS}^2\,,
\end{eqnarray}
We choose $M=1$. Finally, $A(\Psi)$ is determined by the condition that the magnetic field must be smooth when crossing the light-cylinder
\be
\left. \frac{\partial \Psi}{\partial R}\right|_{R=1} =\frac{1}{2}A(1,z)A^{\prime}(1,z)\,.
\label{LC}
\ee
To ensure that our numerical calculation produces accurate results we solve for a force-free pulsar magnetosphere following the procedure described in \cite{Contopoulos:1999} at a resolution of $200$ points in R, equally spaced inside and outside the light-cylinder, and 600 points in $z$. The computational domain extends to $R_{max}=2$ and $z_{max}=6$. We note that increasing the size of the computational domain by a factor of $2$ leads to a difference of less than $1\%$ on the overall result and doubling the resolution changes up to $2\%$ the solution. To account for the return current flowing on the separatrix between the open closed field lines we approximated the $\delta$-function by a Gaussian centered at $\Psi=0.975 \Psi_{0}$ with width $ 0.025/\sqrt{2}$, where $\Psi_{0}$ is the magnetic flux function corresponding to the last open field line. The results are within a $2\%$ agreement with the results of \cite{Gruzinov:2005} (see their Table I). Indeed we find that $\Psi_{0}=1.30$, see run P0 in Table 1 and top left panel on Figure \ref{Figure:2}. 

\begin{table}
	\centering
	\caption{Solutions of force-free magnetospheres confined by an external magnetic field. The columns contain the name of the run, the strength of the external magnetic field with negative being the counter-aligned configurations, the value of $\Psi$ at the last open magnetic field line, the open magnetic flux contained within the light-cylinder and finally the location where the first field line not linked to the pulsar crosses the equator. Note that model C0 corresponds to a non-rotating dipole, where the solution is given by simple addition of the two magnetic fields.}
	\label{tab:Solution}
	\begin{tabular}{ccccl} 
		\hline
		Name & $B_{0}$ & $\Psi_{0}$ & $\Psi_{LC}$ & $l_{0}$   \\
		\hline
        P0 & 0 & 1.30 & 0. & - \\
        A1 & 0.90 & 1.71 & 0.68 & 1.50\\
        A2 & 0.98 & 1.76 & 0.74 & 1.30\\
        A3 & 1.16 & 1.86 & 0.85 & 1.00\\
        A4 & 2.25 & 2.43 & 1.59 & 0.70\\
        A5 & 5.30 & 3.17 & 2.75 & 0.50\\
        A6 & 10.0 & 3.58 & - & 0.43\\
        C0 & -5.30 & 0. & - & 0.72\\
        C1 & -5.30 & 0. & - & 0.88\\
        C2 & -10.0 & 0. & - & 0.62\\
 		\hline
	\end{tabular}
\end{table}

In \cite{Gourgouliatos:2011a}, we sketched the structure of the magnetosphere of a pulsar embedded in an external magnetic field. We found that the qualitative characteristics of these magnetospheres depend on the spin frequency of the pulsar, its magnetic dipole moment and the external magnetic field. Keeping the pulsar dipole strength and the external field constant we found that a slow rotator may not form an equatorial current sheet at all. Moreover, if the fields are counter-aligned the magnetosphere takes the form of a modified Dungey-sphere \citep{Dungey:1961}. Rapid rotators, on the contrary, form equatorial current sheets which are truncated outside the light-cylinder radius. 

Let the external field be $\bm{B}=B_{0}\bm{\hat{z}}$. We have chosen to vary the parameter $B_0$ instead of considering a rapid (slow) rotator as it is essentially equivalent to assume a weak (strong) external field or a strong (weak) dipole field keeping the other two parameters fixed. Essentially this expresses the ratio of the strength of the uniform field compared to the strength of the dipole field at the position of the light-cylinder.

\subsection{Rapid rotators - Weak external field}
\label{RR}
Let us first consider a weak external magnetic field $B_{0}$. This field dominates at $R\to \infty$, whereas the pulsar dipole will dominate near the origin. An equatorial current sheet will form beyond the light-cylinder up to $l_{0}$, where, in analogy with the model problem, it  splits into two branches flowing on the separatrix between the field lines which are connected to the pulsar and the ones that are not. The equatorial part of the current sheet carries both poloidal and toroidal currents, whereas the two branches only have poloidal currents. This is because of the discontinuities of the magnetic field. On the equator both the poloidal and toroidal components of the magnetic field are discontinuous. On the branches the poloidal field is continuous while the toroidal field is not. This is because the open field lines that are connected to the star have a toroidal component, whereas the ones beyond the separatrix are potential without any toroidal field.

We split the domain in four  regions satisfying different equations.  Region I contains closed magnetic field lines within the light-cylinder, they are connected to the pulsar, corotate and have no toroidal magnetic field. The flux function in this region satisfies $\Psi_{\rm I}(R,z)>\Psi(1,0)$ and $R<1$. Region IIa contains open magnetic field lines that are linked to the pulsar have a toroidal field, corotate and cross the light-cylinder. The flux function satisfies $\Psi(1,z\to \infty)<\Psi_{\rm IIa} (R,z)< \Psi(1,0)$. Region IIb contains open magnetic field lines that are connected to the pulsar, corotate, have a toroidal magnetic field but do not cross the light-cylinder. The flux function satisfies $\Psi_{\rm IIb} (R,z)<\Psi(1,z\to \infty)$. Finally, region III contains magnetic field lines not connected to the pulsar, without any toroidal field which are not corotating. The field there is potential and the magnetic flux function satsfies $\Psi_{\rm III}(R,z)>\Psi(1,0)$ and $R>1$. One can see that the magnetic flux in regions I and III may have the same value but they distinguished by the fact that the former lies within the light-cylinder and the latter outside.

The magnetic field satisfies the following equations in each region
\begin{eqnarray}
(1-R^2)\left(\frac{\partial^2 \Psi}{\partial R^2}-\frac{1}{R}\frac{\partial \Psi}{\partial R}+\frac{\partial^2 \Psi}{\partial z^2}\right)-2R\frac{\partial \Psi}{\partial R} =0\,,\,& {\rm (I)} \label{RI}\\
(1-R^2)\left(\frac{\partial^2 \Psi}{\partial R^2}-\frac{1}{R}\frac{\partial \Psi}{\partial R}+\frac{\partial^2 \Psi}{\partial z^2}\right)-2R\frac{\partial \Psi}{\partial R} =- AA^{\prime}\,,\,& {\rm (IIa,~ IIb)}\label{RII}\\
\frac{\partial^2 \Psi}{\partial R^2}-\frac{1}{R}\frac{\partial \Psi}{\partial R}+\frac{\partial^2 \Psi}{\partial z^2}=0\,,\,& {\rm (III)}\,.\label{RIII}
\end{eqnarray}
Next we consider the boundary conditions. Near the origin $R^2+z^2\to 0 $ the magnetic field is a dipole. Thus, we set on the surface of the star ($r^2_{NS}=R^2+z^2$) the flux function to be $\Psi= R^2/(R^2+z^2)^{3/2}$. On the $z-$axis it holds that  $B_{R}(0, z)=0$, which in terms of $\Psi$ becomes $\Psi(0,z)=0$. Here, we have used the fact that $\Psi$ is determined up to an additive constant, which is set equal to zero. On the equator and while inside the light-cylinder it is $B_R(R<1, 0)=0$, therefore $\partial \Psi/\partial z |_{R<1,\, z=0}=0$; outside the light-cylinder and in the region of the equatorial current sheet it is $B_{z}(1<R<l_0,0)=0$ thus $\Psi(1<R<l_0,0)=\Psi(1,0)=\Psi_{0}$ which corresponds to the last open field line. Beyond the separatrix, the field on the equator is $B_R(R>l_0, 0)=0$, thus $\partial \Psi/\partial z |_{R>1,\, z=0}=0$. At large distances the external field is expected to dominate. In particular at large axial distance ($R\to \infty$) it is $B_{z}=B_{0}$ and $B_{R}=0$. The situation is somewhat more complicated for the field lines of regions IIa and IIb at $z\to \infty$. Setting $B_{z}=B_{0}$ does not correspond to a solution unless a special choice of $A=A(\Psi)$ is made. To allow for a more general solution, we demand that $B_{R}(R, z\to\infty)=0$. Since, in practice we are simulating a finite rectangular box in $R\in [0, R_{\rm max}]$ and $z\in [0, z_{max}]$, thus the actual conditions employed at the boundaries of the computational box are $\Psi(R_{\rm max}, z)=\Psi_{\rm max}$ and $\partial \Psi/\partial z |_{R, z_{\rm max}}=0$, where $\Psi_{\rm max}=B_{0}R^2_{max}/2$. 

In the isolated pulsar magnetosphere solution, the form of $A=A(\Psi)$ is determined by the demand that the magnetic field lines cross smoothly the light-cylinder. This is used for the field lines of region IIa, however, the magnetic field lines of region IIb do not cross the light-cylinder, thus we cannot make use of equation \ref{LC} to determine $A$. In principle, one can make a choice of $A(\Psi)$ and then find the corresponding solution. In our approach we make the choice that the poloidal magnetic field in this region is equal to the externally imposed magnetic field $B_{0}$. For this to hold, the flux function needs to be $\Psi(R<1, z\to \infty)=B_{0} R^2/2$. Substituting this expression into \ref{RII} we obtain $A A^{\prime}_{\rm IIb}=2B_{0} R^2=4\Psi$. This choice for $AA^{\prime}$ is then used to solve the pulsar equation in region IIb.

To integrate the pulsar equation we modify the procedure of simultaneous relaxation used in \cite{Contopoulos:1999}. We start from an initial trial for $\Psi$ and $A(\Psi)$ and use the Gauss-Sidel algorithm \citep{Press:1992} to find the corresponding solution. We use the values on either side of the light-cylinder, taking the average, to correct our choice of $AA^{\prime}(\Psi)$. Then the new expression is used to integrate the equation until convergence is achieved. Similarly we update $AA^{\prime}_{\rm IIb}$ when we solve in region IIb. In these simulations we need to make an initial guess for the extent of the equatorial current sheet subject to the external uniform field. To do so we run several simulations with combinations of current sheet lengths and external magnetic fields until convergence is achieved. Varying the value of the external magnetic field, it changes the length of the current sheet as well in accordance to the analytical solution of the model problem: stronger magnetic fields truncate the current sheet closer to the light-cylinder. We have solved for  equilibria corresponding to three choices of external magnetic field leading to current sheets extending to $1.5R_{LC}$, $1.3R_{LC}$ and the limiting case where the current sheet completely disappears (Figure \ref{Figure:2} , top row second and third panel, and bottom row first panel).  We find that the last open field line has a higher value of $\Psi_{0}$, thus a larger fraction of magnetic flux is in the open field line lines, see runs P1, P2, P3 in Table 1.

Reversing the direction of the external magnetic field does not affect the pressure and tension of the magnetic field lines, therefore, the solutions still hold. Thus, this solutions corresponds both to an aligned and to a counter-aligned rotator. What does change though, is the current sheets forming on the separatrix between regions IIa an III: in the case of the alinged rotator the current flowing on the sheet is only poloidal due to the discontinuity of the $B_{\phi}$ component, whereas in the counteraligned case there is an an extra toroidal component due to the discontinuity of the poloidal magnetic field.

\subsection{Slow rotators - Strong external field}

In the slow rotator regime the separatrix crosses the equatorial plane within the light-cylinder. Thus, no equatorial current sheet forms. Similarly to  rapid rotators we can distinguish four  regions. Region I as in the rapid rotator has fields lines whose both ends are connected to the neutron star and cross the equatorial plane. Regions IIa and IIb contain field lines that emerge from the neutron star and extend to infinity, which either cross the light-cylinder (IIa) or not (IIb). Finally, region III, contains field lines that are not linked to the neutron star at all. The main differences between rapid and slow rotators are as follows. Firstly, the boundaries separating regions I from IIa, and IIa from III intersect on the equator at distance $l_{0}$ from the origin, where $l_{0}<1$. Thus no equatorial current sheet forms. Secondly,  very slow rotators do not have region IIb, as illustrated in solution A6 (see Table 1 and Figure \ref{Figure:2}). 

The equations corresponding to slow rotators are the same as the ones of the rapid rotator \ref{RI}-\ref{RIII} and we follow the same procedure for the determination of $A=A(\Psi)$ demanding the field lines of region IIa to cross smoothly the light-cylinder and using a linear relation between $AA^{\prime}=4\Psi$ for the solution in IIb. With respect to the boundary conditions, we impose the same boundary conditions at $R=0$, $R=R_{\rm max}$, $z=z_{\rm max}$ and the surface of the star are identical to those of rapid rotators. However, due to the absence of an equatorial current sheet, the boundary condition at $z=0$ becomes $B_{R}(R, 0)=0$, so that the magnetic field crosses the equator perpendicularly. Subject to these constrains we integrate the differential equations and we find the structure of the magnetosphere as shown in Figure \ref{Figure:2} bottom row, where the equatorial current sheet has been replaced by a null point. In the solutions presented, the last open field line crosses the equator at $0.7R_{LC}$, $0.53 R_{LC}$ and $0.42R_{LC}$ (Table 1, runs A4, A5, A6). Stronger uniform magnetic fields push this point closer to the star and more magnetic flux is carried by the open magnetic field lines. 

Therefore, for an aligned rotator the overall magnetic field structure can be described through a continuum of states which depend on the magnetic field strength (or the spin frequency). As the external magnetic field decreases the transition between the field lines connected to the pulsar and the rest occurs at greater distances from the pulsar. A qualitative transition occurs when this happens outside the light-cylinder, leading to the formation of a finite current sheet instead of a null point. 

\subsection{Rotating Dungey-sphere}
\label{DS}

\begin{figure*}
\includegraphics[width=0.68\columnwidth]{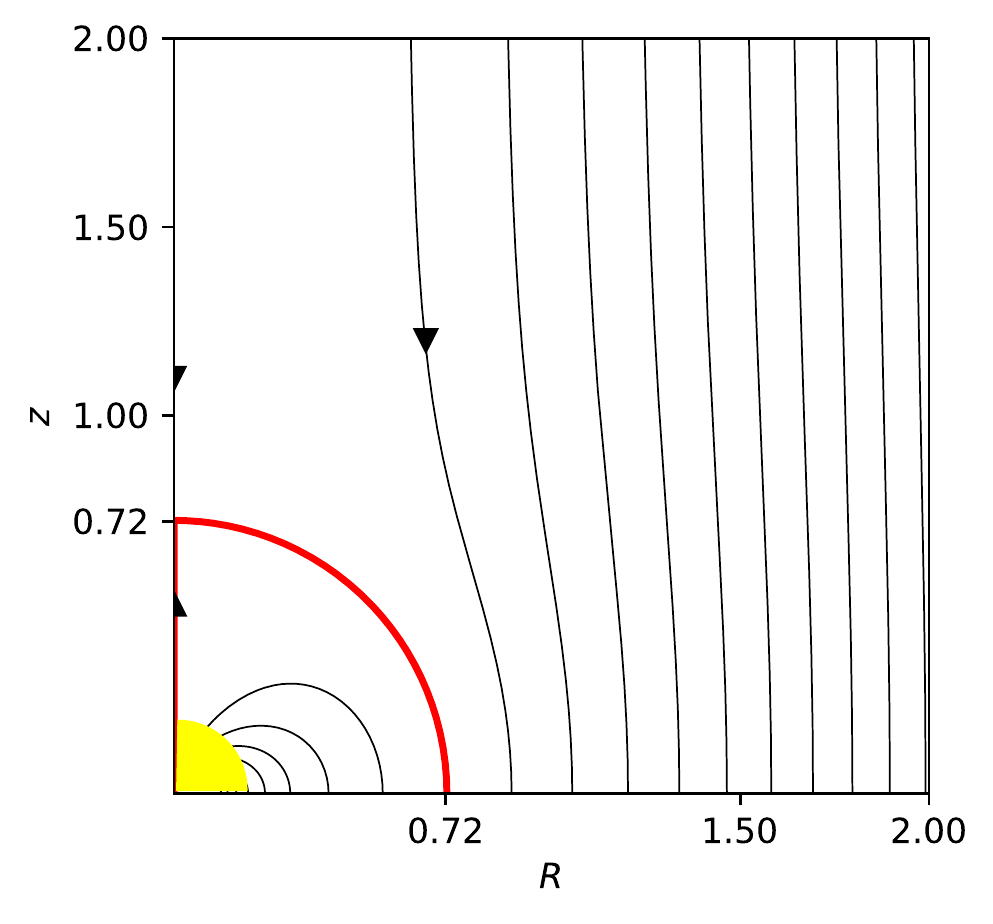}
\includegraphics[width=0.68\columnwidth]{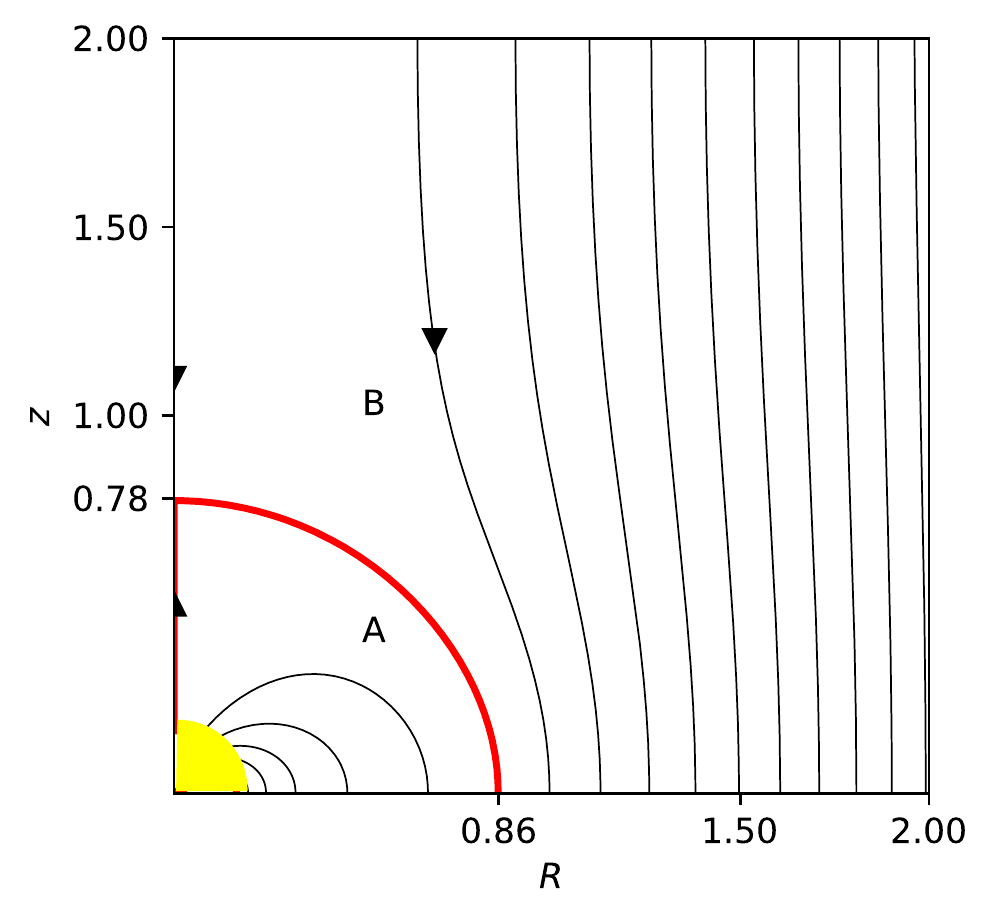}
\includegraphics[width=0.68\columnwidth]{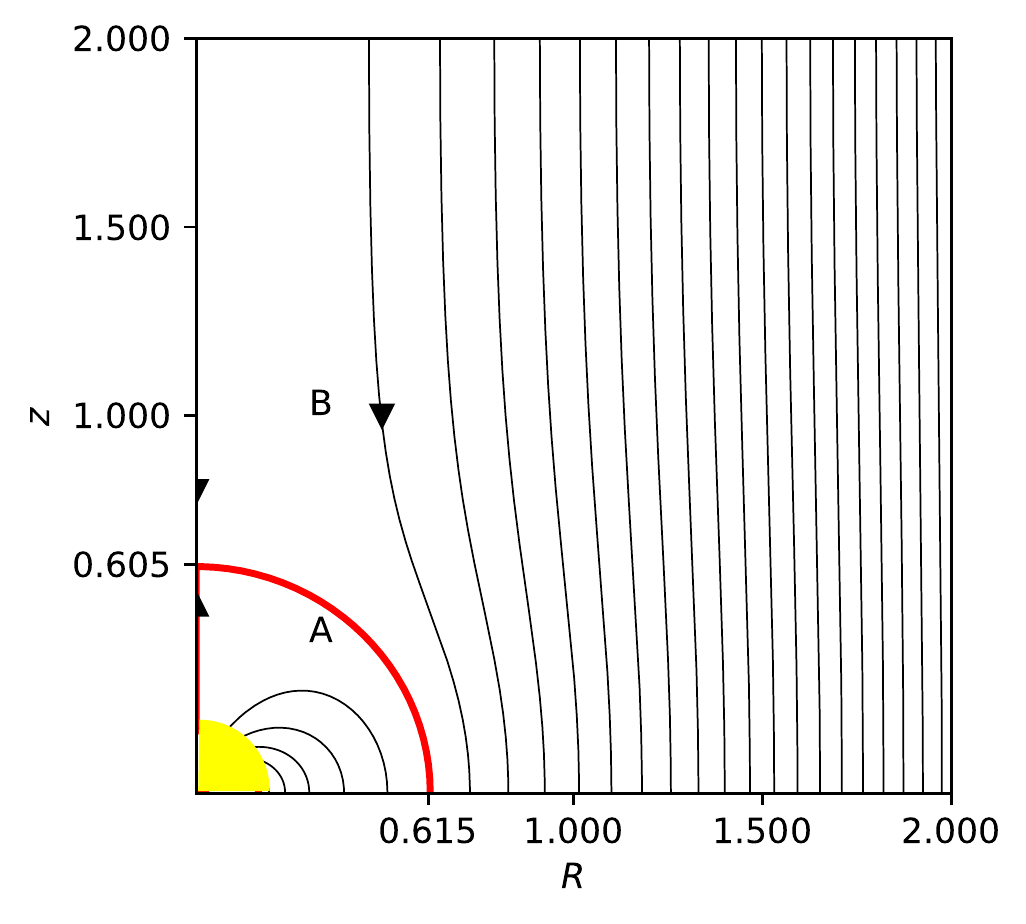}
\caption{Magnetic field structure for the Dungey-sphere solutions with a dipole field and a counteraligned uniform field.  The field lines connected to the star are within the red curve. Left panel: Model C0, a superposition of a non-rotating dipole magnetic field with an an external uniform magnetic field $B_{0}=-5.3$, the field satisfies everywhere equation \ref{RA}. Middle panel: Model C1, magnetic field moment and external magnetic field as in the previous case, but now the dipole field is rotating at $\Omega=1$ so that the light-cylinder is located at $R=1$, the deformation from a spherical to an oblate boundary is obvious. Right panel: Model C2. Superposition of a magnetic field $B_{0}=-10$, to a rotating  the magnetic field with unit magnetic moment, the light-cylinder is located at $R=1$. The deviation from a sphere in this case is minimal. }
\label{Figure:3}
\end{figure*}

A qualitatively different configuration can occur when the magnetic field of the neutron star and the external field are counter-aligned. This setup leads to the so-called Dungey-sphere solution. If the magnetic dipole is not rotating we can simply add the two magnetic fields. Consider a dipole with magnetic flux function $\Psi_{d}=MR^{2}/(R^2+z^2)^{3/2}$ and the uniform field is ${\bf B}=B_{0}\hat{\bf z}$, with $M>0$ and $B_{0}<0$, the flux function describing the superposition of the two fields is $\Psi=MR^{2}/(R^2+z^2)^{3/2}+\frac{1}{2}B_{0}R^2$. The field lines that are connected to the dipole have all $\Psi>0$ and lie within a sphere of radius $R_{0}=(-2M/B_{0})^{1/3}$. If the dipole rotates the induced electric field in the corotating region would affect the overall equilibrium. The configuration is described by the following equations: 
\begin{eqnarray}
(1-R^2)\left(\frac{\partial^2 \Psi}{\partial R^2}-\frac{1}{R}\frac{\partial \Psi}{\partial R}+\frac{\partial^2 \Psi}{\partial z^2}\right)-2R\frac{\partial \Psi}{\partial R} =0\,,\,& {\rm (A)} \label{RA}\\
\frac{\partial^2 \Psi}{\partial R^2}-\frac{1}{R}\frac{\partial \Psi}{\partial R}+\frac{\partial^2 \Psi}{\partial z^2}=0\,,\,& {\rm (B)}\,,\label{RB}
\end{eqnarray}
where the light-cylinder is $R=1$. 
Region A corresponds to the field lines that are connected to the dipole and corotate. These field lines do not have any toroidal field as they are anchored on both hemispheres of the star. Nevertheless, an electric field is induced due to corotation. Region B, contains field lines which are not connected to the dipole. The field there is potential. We have solved for a non-rotating dipole with $M=1$, where an external field $B_{0}=-5.3$ is superposed (C0), and we found that this field is confined within a sphere of radius $0.72$, Figure \ref{Figure:3} left panel. We then solved the same problem with a rotating dipole. In model C1 the confining surface is approximately an oblate spheroid and intersects the equator at $R=0.86$ and  the axis at $z=0.78$, Figure \ref{Figure:3} middle panel. Finally, we considered an external field $B_{0}=-10$, model C2. Here, the confining surface intersects the equator at $R=0.615$ and the axis at $z=0.605$. Weaker external fields brings the last closed field line closer to the light-cylinder until it eventually reaches it. Then, part of the magnetic flux opens and the magnetic field develops a toroidal component, funnelled inside a cylinder round the axis of the dipole. We found that this should happen for magnetic with intensities below $5.1$.  Then, the rotating and the non-rotating part of the flow will be separated by a current sheet, in this case though, there are two discontinuities: one involving the toroidal field which leads to a poloidal current and another involving the poloidal field which reverses direction and leads to a toroidal current. Due to the discontinuity of the poloidal field the numerical scheme employed did not converge to a solution.

An interesting situation occurs by comparing solutions A5 and C1. Since reversing the magnetic field beyond the separatrix in solution A5 does not affect the equilibrium, we find that for the same magnetic field one can have two  solutions with different topologies. One where the magnetic field emerging from the neutron star is fully confined within a sphere (C1) and another where the magnetic field linked to the pulsar extends to infinity (A5-reversed). Solution C1 does not contain a current sheet separating regions A and B, whereas A5-reversed contains a current sheet with both toroidal and poloidal currents flowing on the separatrix. This could make solution A5-reversed prone to resistive instabilities. We have evaluated the energy contained inside a cube whose side is four times the light-cylinder and we have excluded a sphere at the centre whose radius is $0.2$ where the star is located. We find that the open configuration contains more energy compared to the closed, in particular the difference is $2.5 B(R_{LC},0)^2 R_{LC}^3$, where $B(R_{LC},0)$ is the field of the dipole at the equator at a distance equal to its light-cylinder. This implies that C1 is energetically favourable compared to A5-reversed. It is then likely that if a pulsar with an initially open magnetic field configuration, experiences an increasing external magnetic field, at some point to become unstable and adopt the structure of a Dugney sphere. Given the difference in energies, this change in topology will release the excess amount of energy in an explosive event.

\subsection{Asymptotic Behaviour}
\label{ASYMPT}

Strong external magnetic fields push the separatrix well within the light-cylinder of the pulsar. While the field within the separatrix corotates, its velocity is much smaller than the speed of light. Thus, the overall dynamics are governed by the equilibrium of the dipole field with the externally imposed one. Using the flux function $\Psi=R^2/(R^2+z^2)^{3/2}+B_{0}R^2/2$, we can estimate the size of the corotating part of the magnetic field by finding where the separatrix intersects with the equatorial plane. At the point of intersection $R_{0}$ the sign of the $B_{z}$ component changes from negative to positive, thus $\frac{d\Psi}{dR}|_{z=0}=0$, thus $R_{0}=B_{0}^{-1/3}$ and the open flux is $\Psi_{0}=\frac{3}{2}B_{0}^{1/3}$, where $\Psi=1$ corresponds to the flux of a non-rotating dipole at $R=1$, $z=0$. 

For weak external fields the large distance limit is relevant. There, the system resembles a split monopole embedded in a uniform field. We can use the estimate of the model problem, equation \ref{SPLIT_MON}, where  the size of the current sheet scales as $l_{0}\propto B_{0}^{-1/2}$. The open flux though, will be close to the value of $1.30$ which is the asymptotic limit for the open field lines. 

\begin{figure}
\includegraphics[width=\columnwidth]{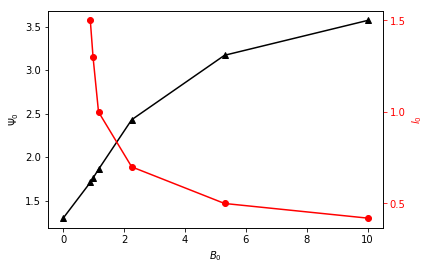}
\caption{Values of $\Psi_{0}$ (black triangles) and $l_{0}$ (red circles) as a function of $B_{0}$ for the numerical solutions found. }
\label{Figure:4}
\end{figure}

These results are consistent with the numerical results in Figure \ref{Figure:4}. Indeed, we notice the increase of the open magnetic flux for stronger magnetic fields and the drop towards the asymptotic value for weaker fields. 

\section{Application to Double Neutron Stars}
\label{Application}
\begin{figure*}
\includegraphics[width=0.98\columnwidth]{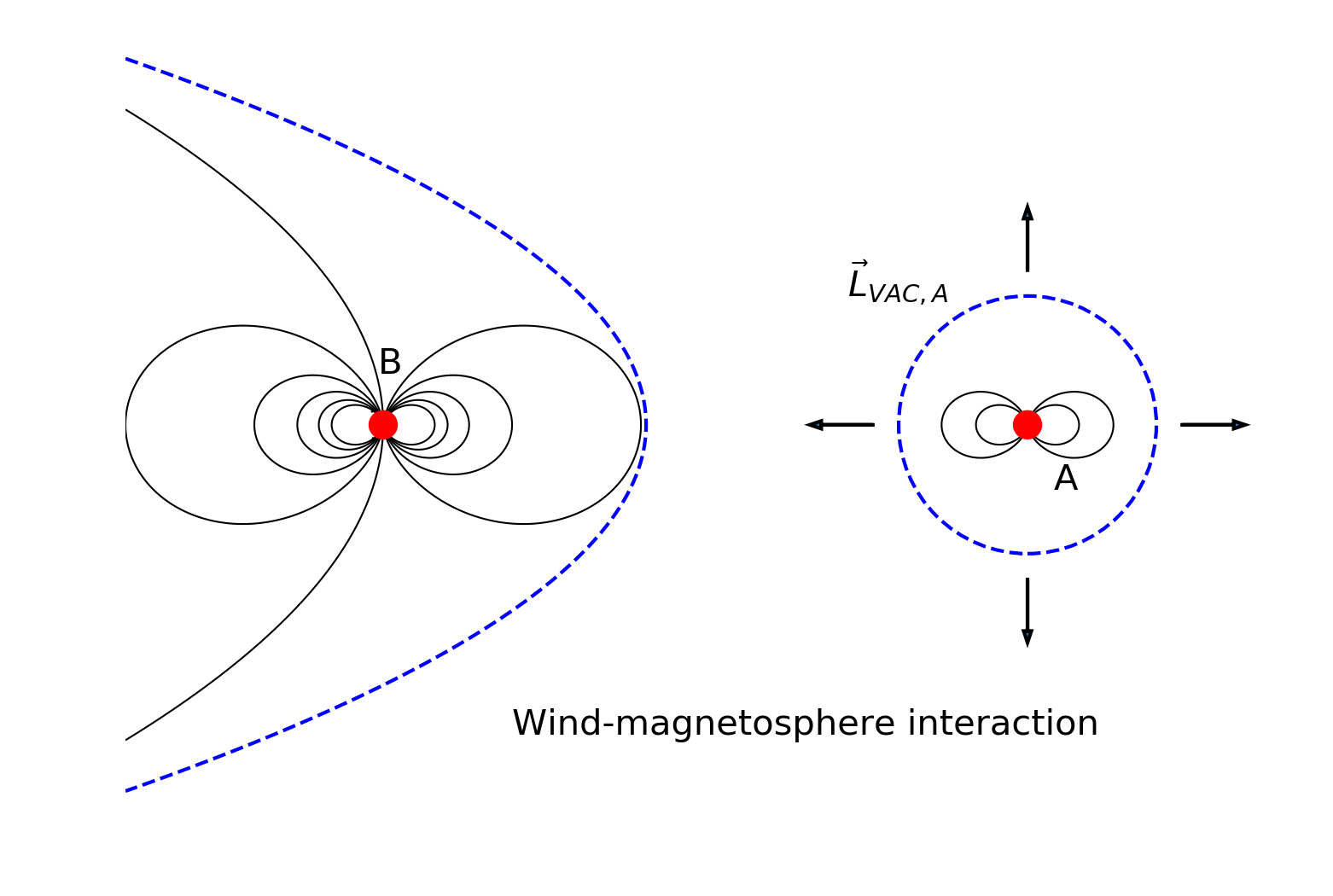}
\includegraphics[width=0.98\columnwidth]{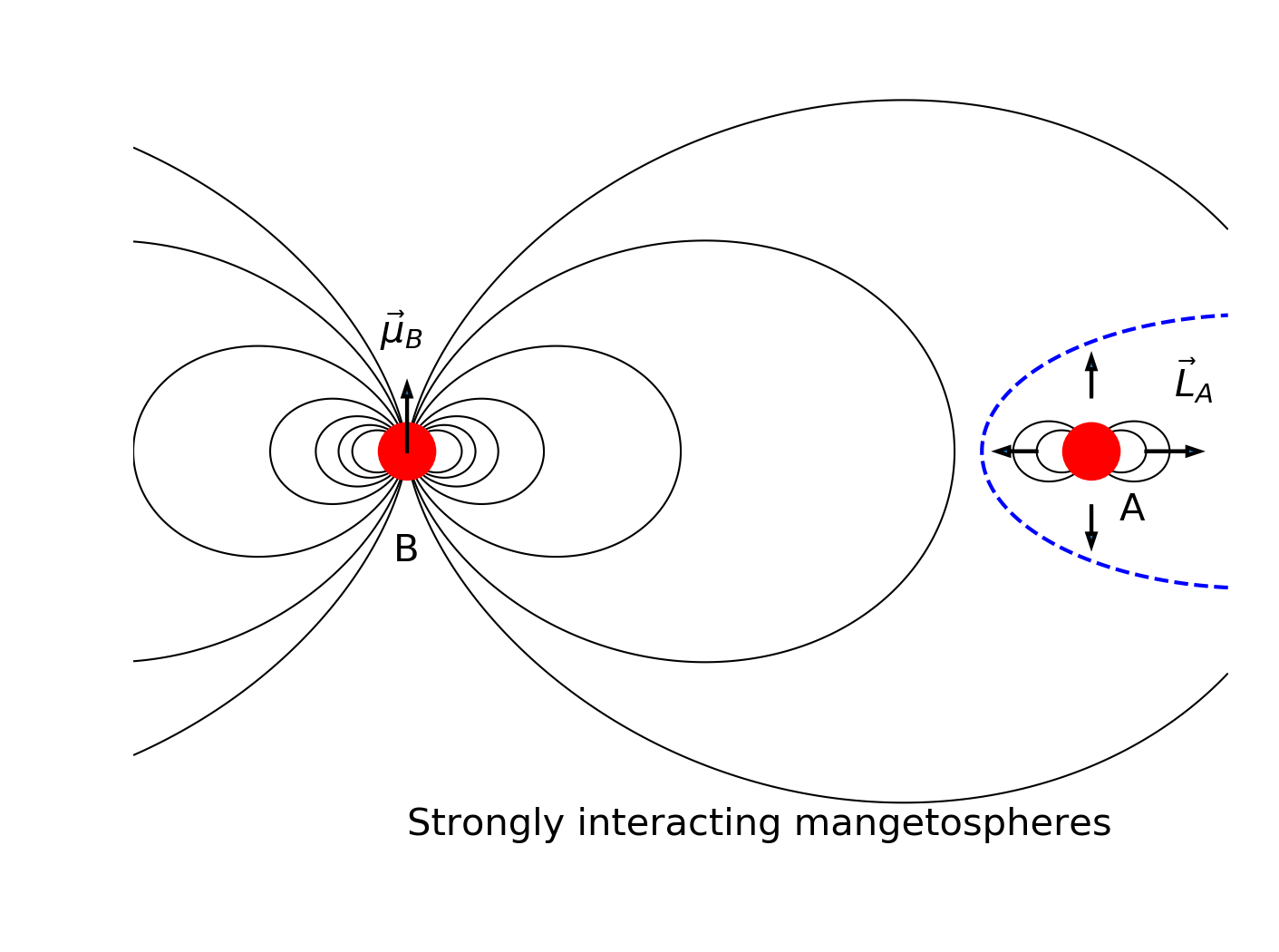}
\includegraphics[width=0.98\columnwidth]{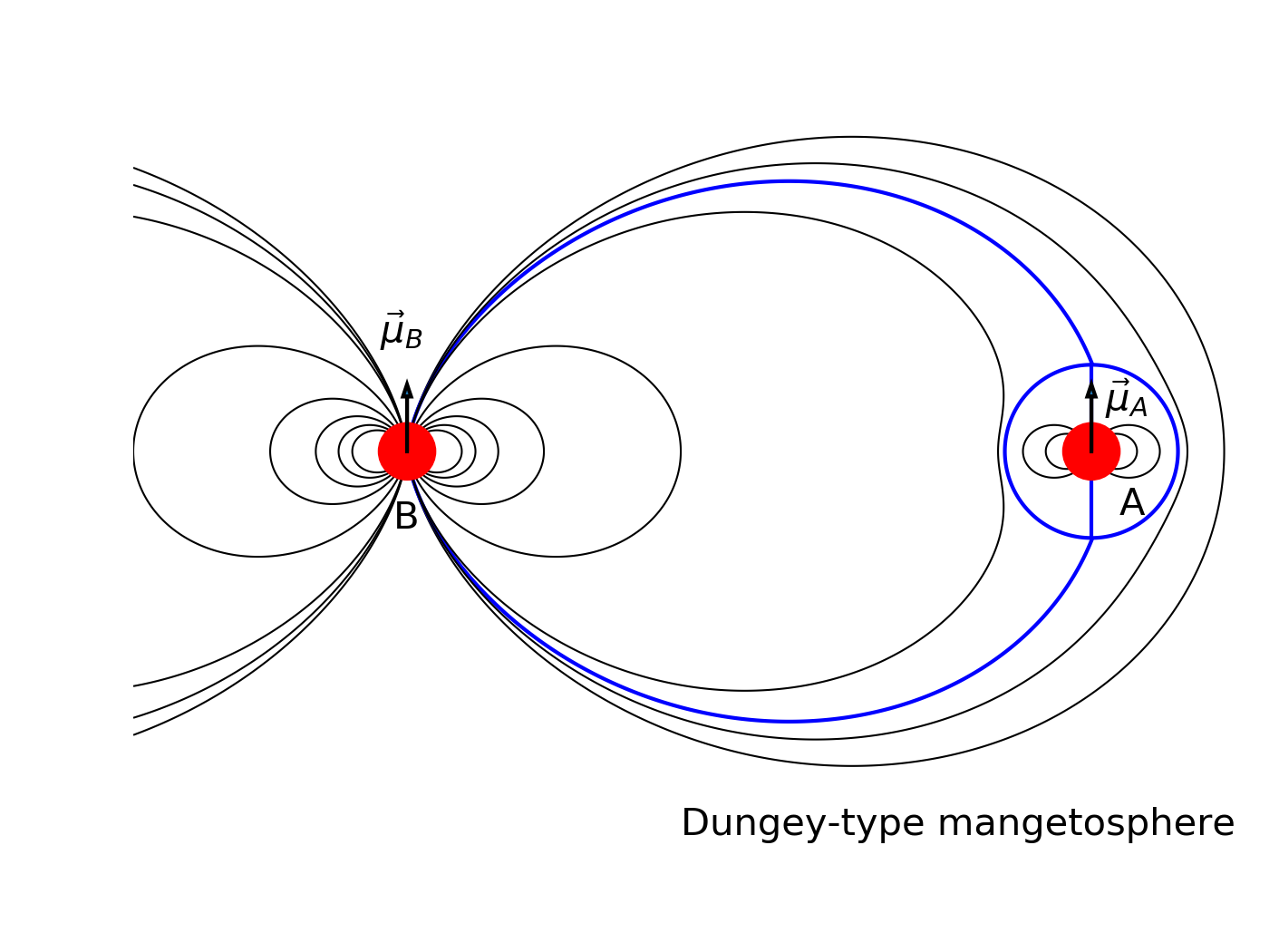}
\includegraphics[width=0.98\columnwidth]{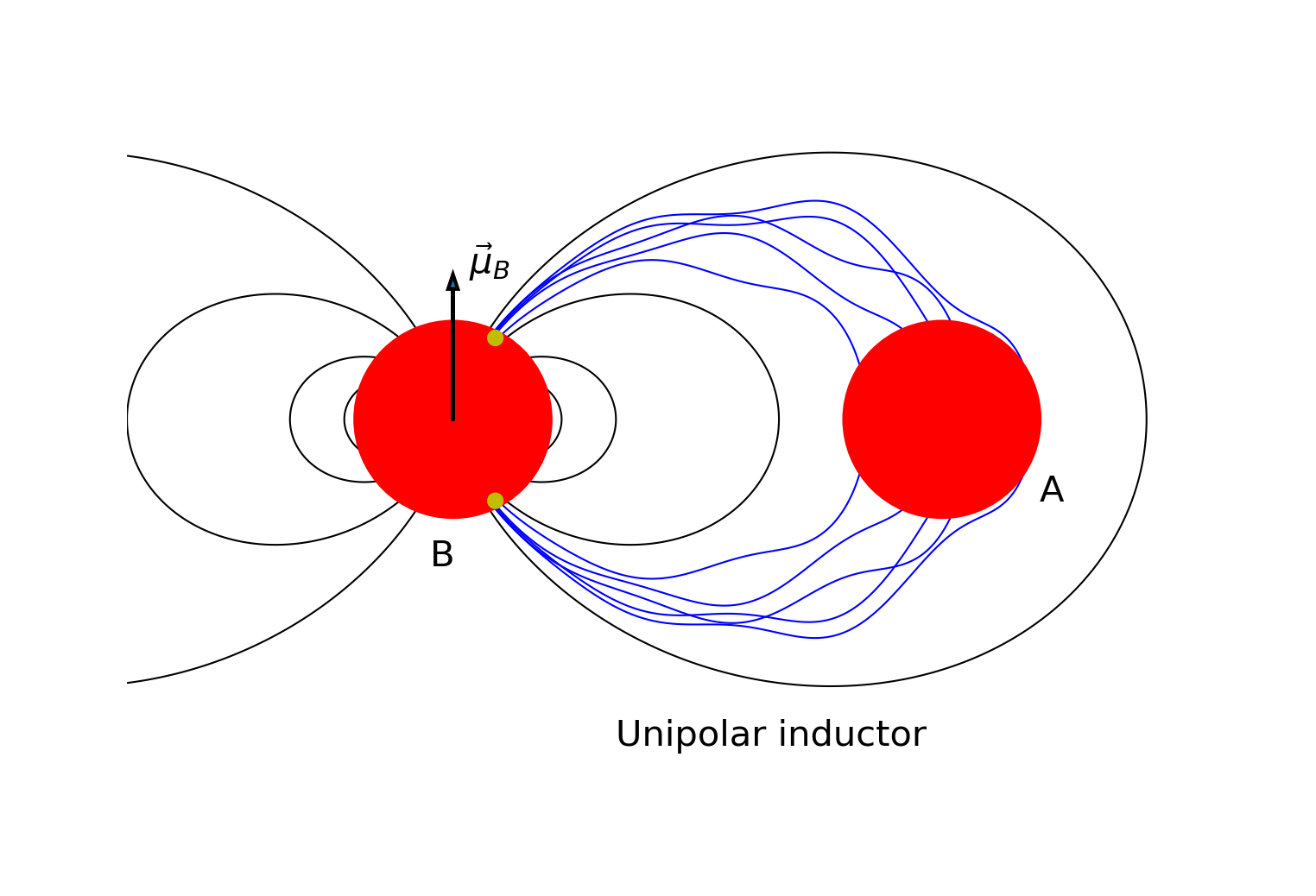}
\caption{Cartoon of the stages of magnetospheric interaction in a DNS with shrinking orbit. At large separations the wind of A may deform the magnetosphere of B (top left), at closer separations, the two magnetospheres will strongly interact with the magnetosphere of A most likely developing a highly anisotropic wind (top right), or being fully confined in a Dungey-sphere configuration (bottom left). Shortly before coalescence, DNSs interact through the unipolar inductor scheme (bottom right). }
\label{Figure:5}
\end{figure*}

Below we explore the implications of these solutions to a DNS consisting of pulsars A and B, as its orbit shrinks due to gravitational wave radiation. Let the two members have equal masses, $1.4M_{\odot}$, and radii, $r_{NS}$, pulsar A a period $P_{A}$, dipole magnetic field $B_{A}$ and pulsar B $P_{B}$ and $B_{B}$ respectively. These magnetic fields correspond to the strengths on their surface at the equator. The distance between the two neutron stars is $d$. Let us further assume that A is a millisecond pulsar with a weaker magnetic field, whereas B is a young pulsar with a stronger magnetic field and slower period, thus $P_{B}\gg P_{A}$ and $B_{B}\gg B_{A}$. 

Once the two pulsars are close enough the wind of pulsar A may affect the magnetosphere of B. At smaller separations the two magnetospheres can become coupled and, depending on the relative orientation of the field, Dungey-sphere configurations may appear. Finally, they come close enough to activate the unipolar inductor mechanism. These stages are schematically shown in Figure \ref{Figure:5}.

\subsection{Wind-Magnetosphere Interaction}
\label{WMI}

Consider the spin-down luminosity of pulsar A, under the simplification of an orthogonal dipole rotator in vacuum
\begin{eqnarray}
L_{VAC,A}=\frac{2}{3 c^{3}}\left(B_{A}r_{NS}^{3}\right)^2\left(\frac{2\pi}{P_{A}}\right)^4\,.
\label{VACUUM}
\end{eqnarray} 
The wind pressure at distance $d_W$ from A is
\begin{eqnarray}
p_{W,A}\approx L_{VAC,A}/(4 \pi d_W^2 c)\,.
\label{WIND-A}
\end{eqnarray} 
The magnetic pressure of B at its light-cylinder is approximately
\begin{eqnarray}
p_{mag,B}\approx \frac{B_{B}^2 r_{NS}^6 }{ 4\pi}\left(\frac{2\pi}{cP_{B}}\right)^6
\label{PRESSURE-B}
\end{eqnarray}
where we have assumed that the magnetic field within the light-cylinder is inversely proportional to the distance from the pulsar to third power. The two pressures become comparable when the separation is 
\begin{eqnarray}
d_{W}\sim \left(\frac{2}{3}\right)^{1/2}\frac{c}{2\pi}\frac{B_{A}}{B_{B}}\frac{P_{B}^3}{P_{A}^{2}}\,.
\end{eqnarray}
Therefore, when the separation between the two pulsars is smaller than $d_{W}$ the magnetosphere of pulsar B will be affected by the wind of A.

\subsection{Strongly interacting magnetospheres}

As the separation of the DNSs decreases, there will be a point at which, the magnetic field of B at the vicinity of A will become stronger than that of A at its light-cylinder.  As shown in the numerical axisymmetric solutions derived in section \ref{NUM_SOL}, a uniform field  comparable or stronger to the pulsar's field at its light-cylinder alters the structure of the magnetosphere of the latter. While, direct application of these solutions to the thee-dimensional time-depended system that arises from such an interaction is a drastic simplification, we can still extract useful conclusions by considering the main properties of these solutions and their asymptotic behaviour.  

Let us assume that pulsar A lies within the light-cylinder of B and approximate the magnetic field of the latter by a magnetic dipole. Its strength at distance $r$ is $B_B(r)=B_B (r_{NS}/r)^3$ and let us further assume that the light-cylinder of the millisecond pulsar is much smaller than their separation, so that the magnetic field of B does not change much in the neighbourhood of A. In the axisymmetric solution we have explored the two extreme cases of aligned and counteraligned pulsars, which set the main framework of the possible configurations. A perfectly aligned magnetic field would lead to drastic enhancement of the open magnetic flux. Using the asymptotic approximation, the open flux scales as $\propto \frac{3}{2} B_{0}^{1/3}$, where $B_{0}$ is the ratio of the magnetic field of pulsar B over the magnetic field of pulsar A at its light-cylinder. Thus the open flux of pulsar A becomes $\Psi_{A} =\frac{3}{2} B_{A}^{2/3}B_{B}^{1/3}r_{NS}^3/d$. Following \cite{Contopoulos:2005} the spin-down power of a pulsar whose open field lines corotate with the neutron star is $L=\frac{2}{3c}\Omega^{2}\Psi_{A}^2$, thus the power emitted by pulsar A when its magnetosphere is confined by the field of pulsar B is 
\begin{eqnarray}
L_{A,~aligned}=\frac{6\pi^2}{c} \frac{B_{A}^{4/3}B_{B}^{2/3}r_{NS}^6}{P_{A}^2 d^2}\,.
\label{Lum}
\end{eqnarray} 
The other limiting case occurs for the counteraliged case, where the field of A is enclosed within an Dungey-type configuration, and the spin-down power is completely shut down, with $L_{A,~Dungey}=0$.

Magnetic field lines emerging from pulsar A could either close onto pulsar A, form a wind, or connect to field lines emerging from pulsar B. Certainly, some of them will be closing onto A, forming the so-called ``dead-zone'', however for a completely enclosed region a highly-symmetric Dungey-type configuration is needed.

Let us next consider the implications of linking field lines from pulsar A to pulsar B. When the two pulsars were sufficiently far (\ref{WMI}), the magnetosphere of pulsar A transitioned to a wind and it was disconnected from that of B. Following their approach, and under ideal-MHD conditions, the field lines maintain their topology and remain disconnected. Allowing for a finite conductivity, field lines from A could, in principle, be connected to those of B. In this case, though, as the two pulsars spin at different rates and orbit each other, these field lines will be highly twisted and wound up. A flux tube linking the two pulsars will be twisted at a rate equal to the $P_{A}^{-1}$. Let its length be comparable to light-cylinder of pulsar B (which is in the same order of magnitude of the separation of the two pulsars) $z_t \sim cP_{B}/(2\pi)$ and its radius $r_t\sim c P_{A}/(2\pi)$ being similar to the light-cylinder of pulsar A. In a manner similar to the magnetic towers \citep{Lynden-Bell:2003}, after $N$ spin periods of A, this process will generate a toroidal field inside the flux tube $b_{\phi,t}\sim \pi N b_{z,t}  P_{A}/P_{B}$, where $b_{\phi,t}$ and $b_{z,t}$ is the axial and toroidal field inside the tube. Taking their ratio to be unity $b_{\phi,t}/b_{z,t}\sim \pi N P_{A}/P_{B} \sim 1$, implies that the system will become unstable within after $N=P_{B}/P_{A}$ rotations. At this point it will exceed the Kruskal-Shafranov limit \citep{Kruskal:1958, Shafranov:1958} and  become prone to kink-instability. Assuming periods of $P_{A}=0.01$s and $P_{B}=1$s this will happen within a second. Thus even if the field lines somehow become connected they would not maintain such a configuration long enough.

Another possibility is that pulsar A forms a wind within the magnetosphere of B. To explore the plausibility of scenario we need first to assess whether the wind pressure emanating from pulsar A is comparable or higher than the magnetic pressure of B at the neighbourhood of the former.  The power of the wind will depend on the open magnetic flux emanating from pulsar A. The vacuum magnetosphere formula in equations \ref{VACUUM} is a good approximation within a factor of a few when the strength of B is comparable to that of A at the light cylinder of the latter. Then, using the expression for the wind pressure from equation \ref{WIND-A} we find that the pressure corresponding to an wind at a distance equal to few light-cylinder of A, is comparable to that of the magnetic pressure of B. Therefore, it is likely that a wind-inflated bubble could form due to the radiation of pulsar A. If such a wind appears it will be highly anisotropic and probably extend in the direction opposite from pulsar B due to the lower pressure there, Figure \ref{Figure:5} top right panel.   

Finally, we remark that as pulsar A moves with respect to the corotating plasma with a relative velocity $\bm{v}$, it will experience an induced $\bm{E}=-\bm{v} \times \bm{B}/c$. This electric field will be in the direction normal to the magnetic field, thus, even if the external magnetic field were aligned with the pulsar dipole, the presence of $\bm{E}$ will break axial symmetry. The complete inclusion of the electric field requires a three dimensional, time-dependent  model which is beyond the scope of this work.

\subsection{Unipolar inductors}

Shortly before coalescence, the two neutron stars will come close enough so that the magnetic field of one component entirely dominates the system. The condition for this to is happen is that the magnetic field of B on the surface of A is higher than that of A. Thus, when the distance between the two pulsars is smaller than
\begin{eqnarray}
d_u\sim \left(\frac{B_B}{B_A}\right)^{1/3}r_{NS}\,,
\end{eqnarray}
the system can transition to the unipolar inductor configuration. Should this be the case, the magnetic field of A will be entirely suppressed and it will act as a unipolar inductor within the magnetosphere of pulsar B \citep{Goldreich:1969b}. Such a configuration will be a precursor of the gravitational wave event and could operate seconds prior to the merger \citep{Hansen:2001, Metzger:2016}.

\section{Discussion}

\begin{table}
	\centering
	\caption{DNS and candidate sources. The columns are: pulsar name, magnetic field strength of the recycled millisecond pulsar (A) and the young pulsar (B), periods of the two pulsars and the gravitational decay times. Decay times longer than 50~Gyr are denoted as $\infty$. The full set of parameters is known only for double pulsar J0737+3039A/B. The parameters from the first four systems of the table were used in Figures \protect\ref{Figure:6} and \protect\ref{Figure:7}. The data for J1946+2052 are from \protect\cite{Stovall:2018}. The data for all other systems were taken from  \protect\cite{Tauris:2017}  Table 1 and references therein. Sources marked with ${\mbox{*}}$ may have companion white dwarfs. }
	\label{tab:Pulsars}
	\begin{tabular}{lccccc} 
			\hline
		Pulsar Name & $B_{A}$ & $B_{B} $  &$P_{A}$ &$P_{B}$& $\tau_{GW}$ \\
		 & $(10^{9}$G) & $(10^{9}$G)  &(ms) &(ms)& (Myr)  \\
		\hline
        J0737-3039 & 6.4 & 1590 & 22.7 &2773.5 & 86\\
        J1946+2052 & 4.0 & - & 17 & - & 46\\
        J1913+1102 & 2.12& - & 27.3 & - & 480\\
        B1913+16 & 22.8 & - & 59.0 & - & 301 \\
        \hline
         J0453+1559 & 3.0 & - & 45.8 & - &2730 \\
         J1518+4904 & 1.07 & - & 40.9 & -& $\infty$ \\
         B1534+12 & 9.7 & - & 37.9 & -& $\infty$ \\
        J1753-2240 & 9.7 & - & 95.1 &-&  $\infty$ \\
        J1755-2550$^{\mbox{*}}$ & - & 880 & - &315.2&  $\infty$ \\
        J1756-2251 & 5.5 & - & 28.5 &-&  1660 \\
        J1811-1736 & 9.8 & - & 104.2 &-&  $\infty$ \\
        J1829+2456 & 1.48 & - & 41.0 &-&  $\infty$ \\
         J1906+0746$^{\mbox{*}}$ & - & 1730 & - &144.1&  309 \\
         J1930-1852 & 58.5 & - & 185.5 &-&  $\infty$  \\
         J1807-2500B$^{\mbox{*}}$ & 0.59 & - & 4.19 &-&  $\infty$  \\
         B2127+11C & 12.5& - & 30.5 &-&  217  \\
 		\hline
	\end{tabular}
\end{table}
\begin{figure*}
\includegraphics[width=0.98\columnwidth]{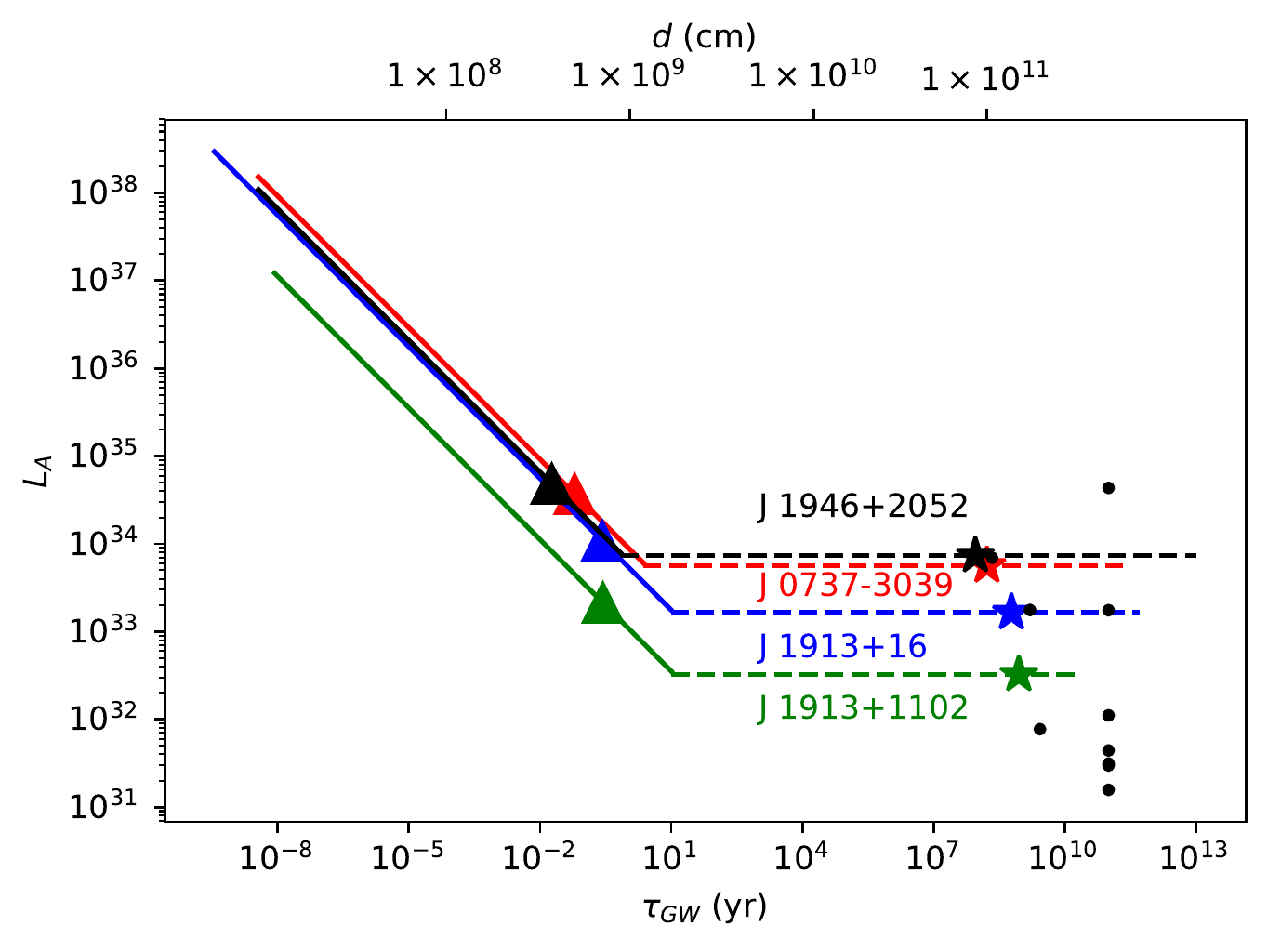}
\includegraphics[width=0.98\columnwidth]{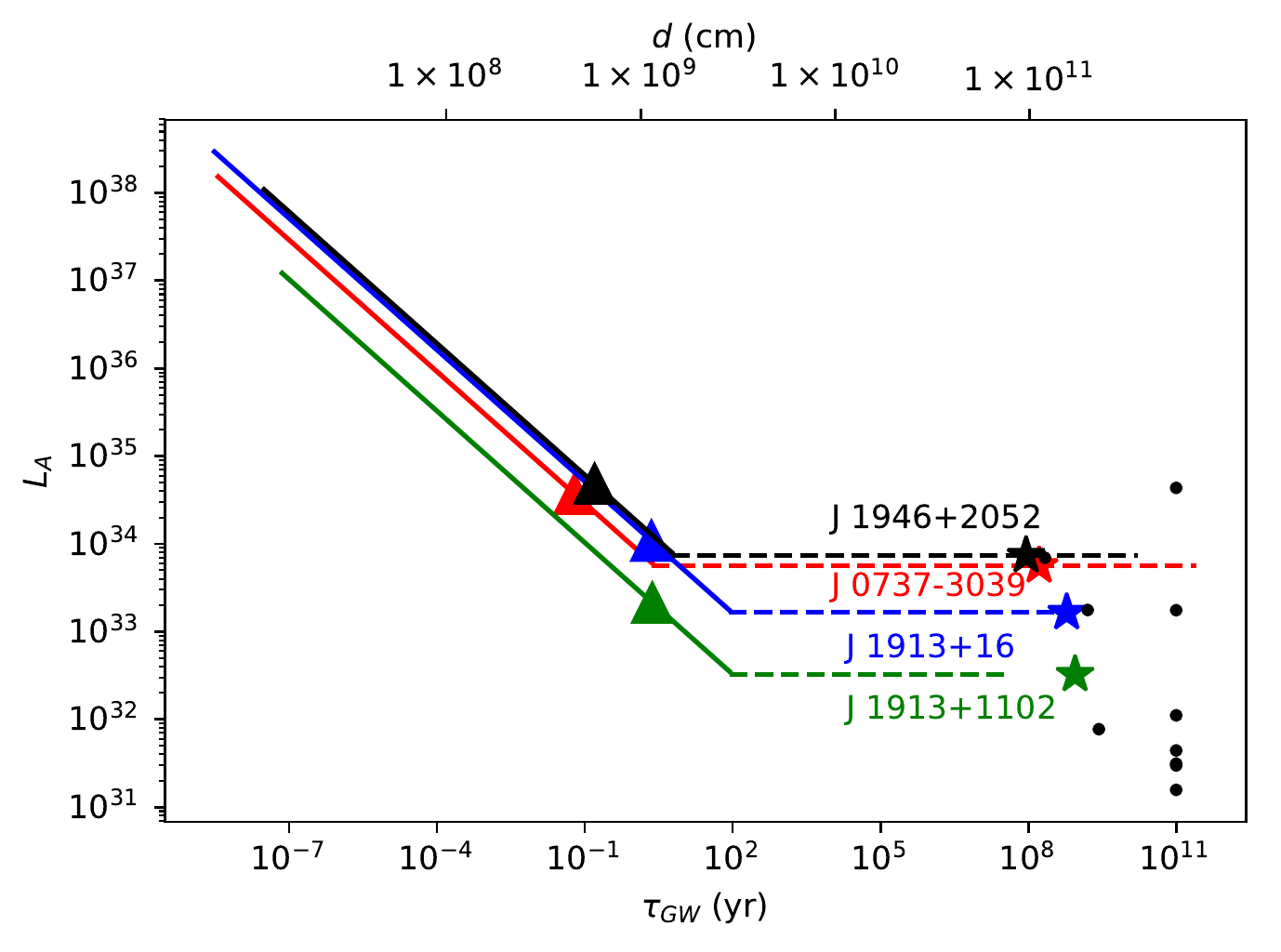}
\includegraphics[width=0.98\columnwidth]{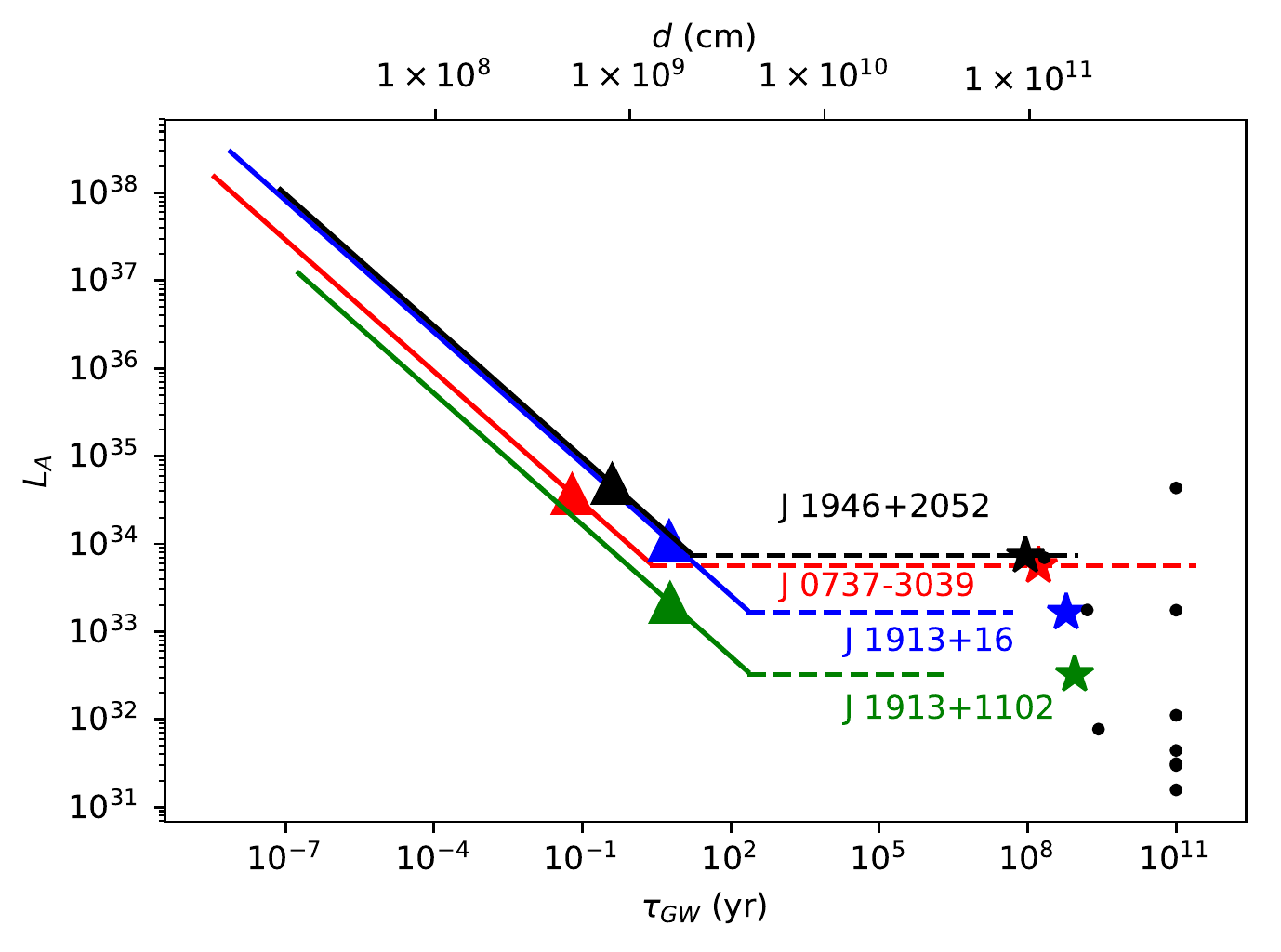}
\includegraphics[width=0.98\columnwidth]{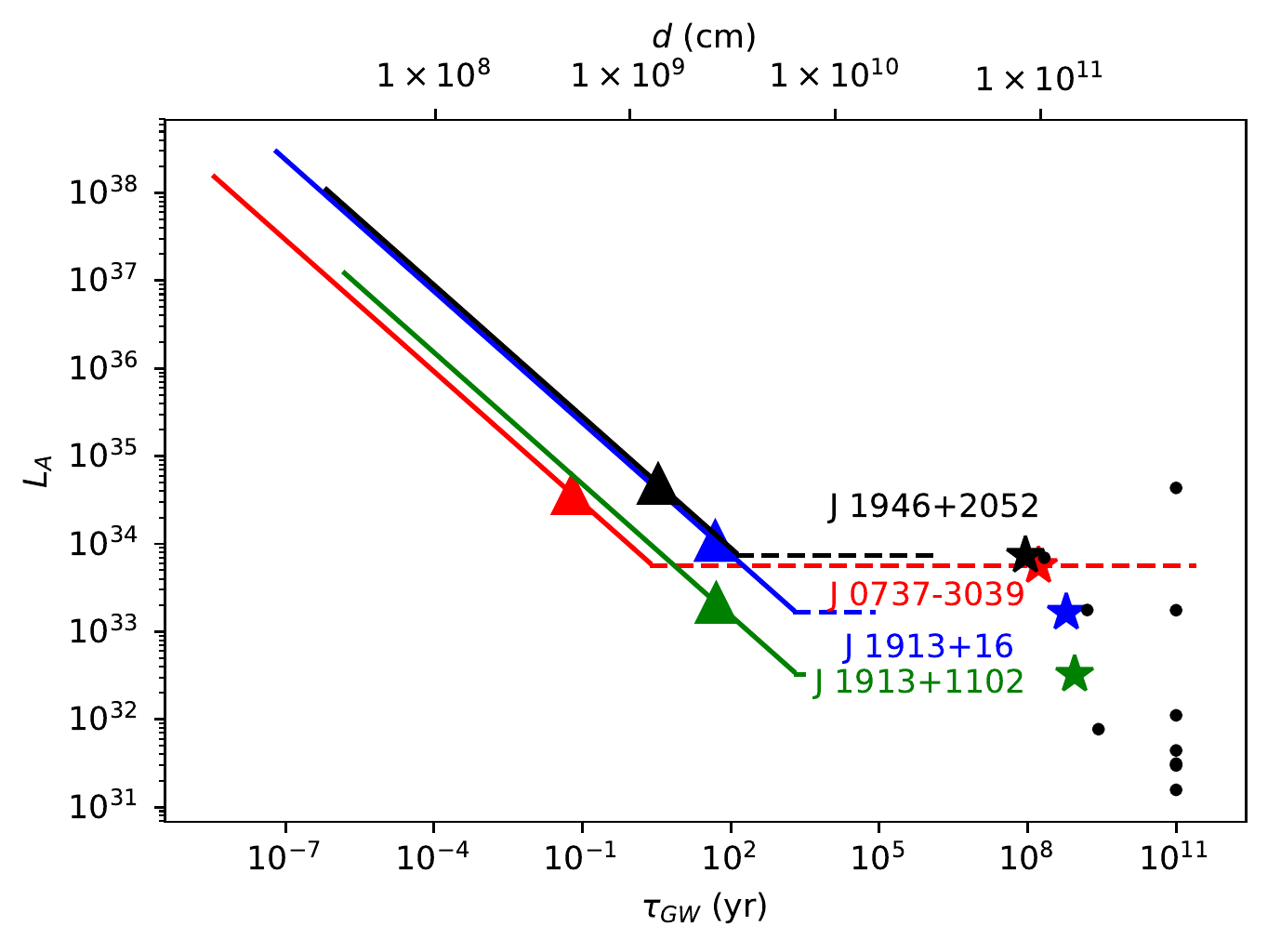}
\caption{The solid line is the enhanced spin-down power of pulsar A as a function of pulsar separation in the strongly interacting magnetosphere case. The dashed line is the spin-down luminosity during wind-magnetosphere interaction, as this phase last less than the characteristic age of pulsar A, its period and period derivative do not change and the spin-down luminosity remains constant. The triangles indicate the maximum separation where the Dungey-sphere configuration is feasible. The red line corresponds to the millisecond member of the double pulsar J0373-3039A. The black, blue and green lines to systems where pulsar A has the properties of J1946+2052, J1913+16 and J1913+1102, while pulsar B is assumed to have a period of $3$~s and a magnetic field of $10^{12}$~G, $5\times 10^{12}$~G, $10^{13}$~G and $5\times 10^{13}$~G, in the top left, top right, bottom left and bottom right panels respectively.  The coloured stars correspond to spin-down power of the millisecond pulsars of the four sources used above. The black dots correspond to the other millisecond pulsars shown in Table \protect\ref{tab:Pulsars}. }
\label{Figure:6}
\end{figure*}
\begin{figure*}
\includegraphics[width=0.98\columnwidth]{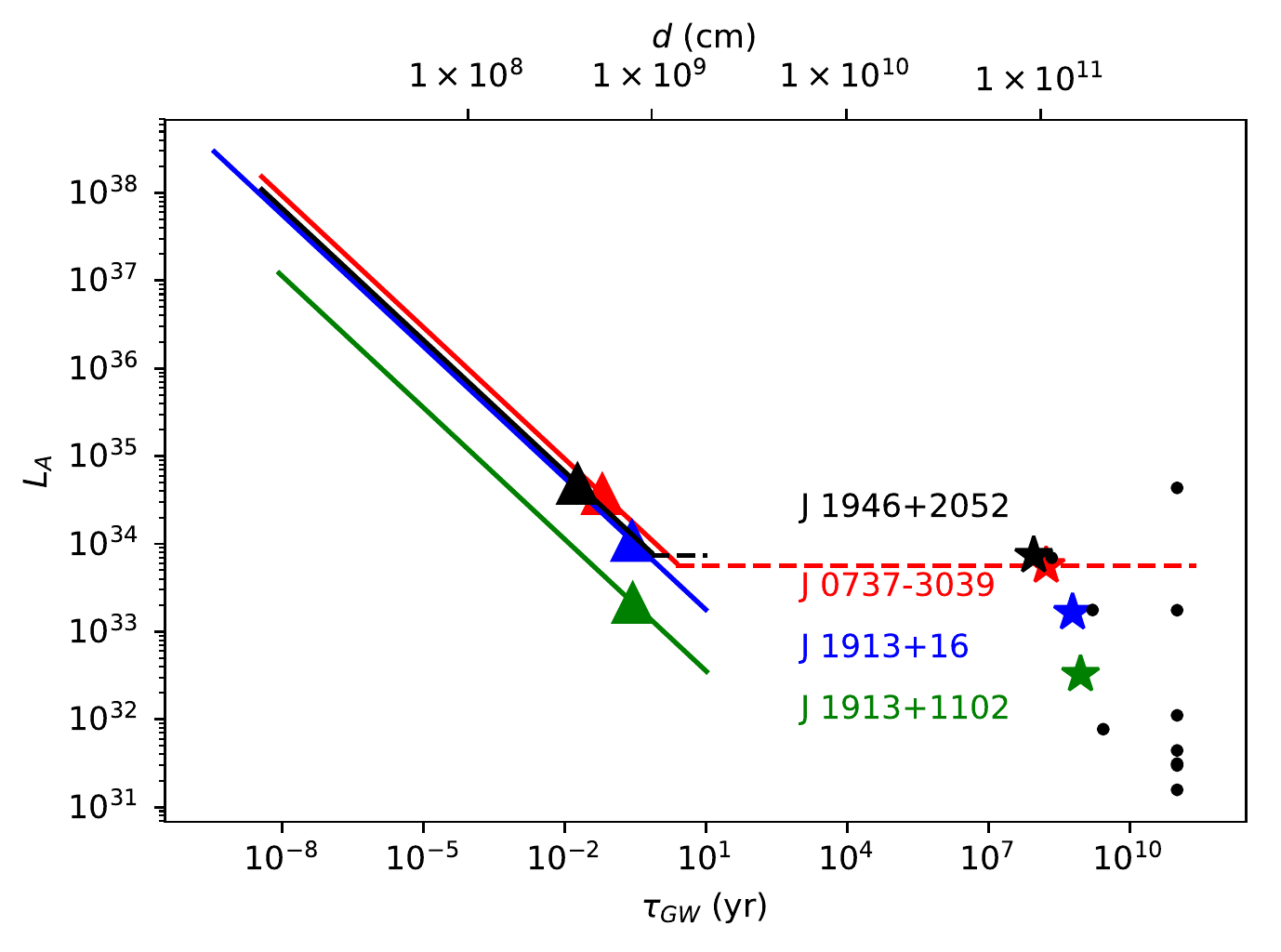}
\includegraphics[width=0.98\columnwidth]{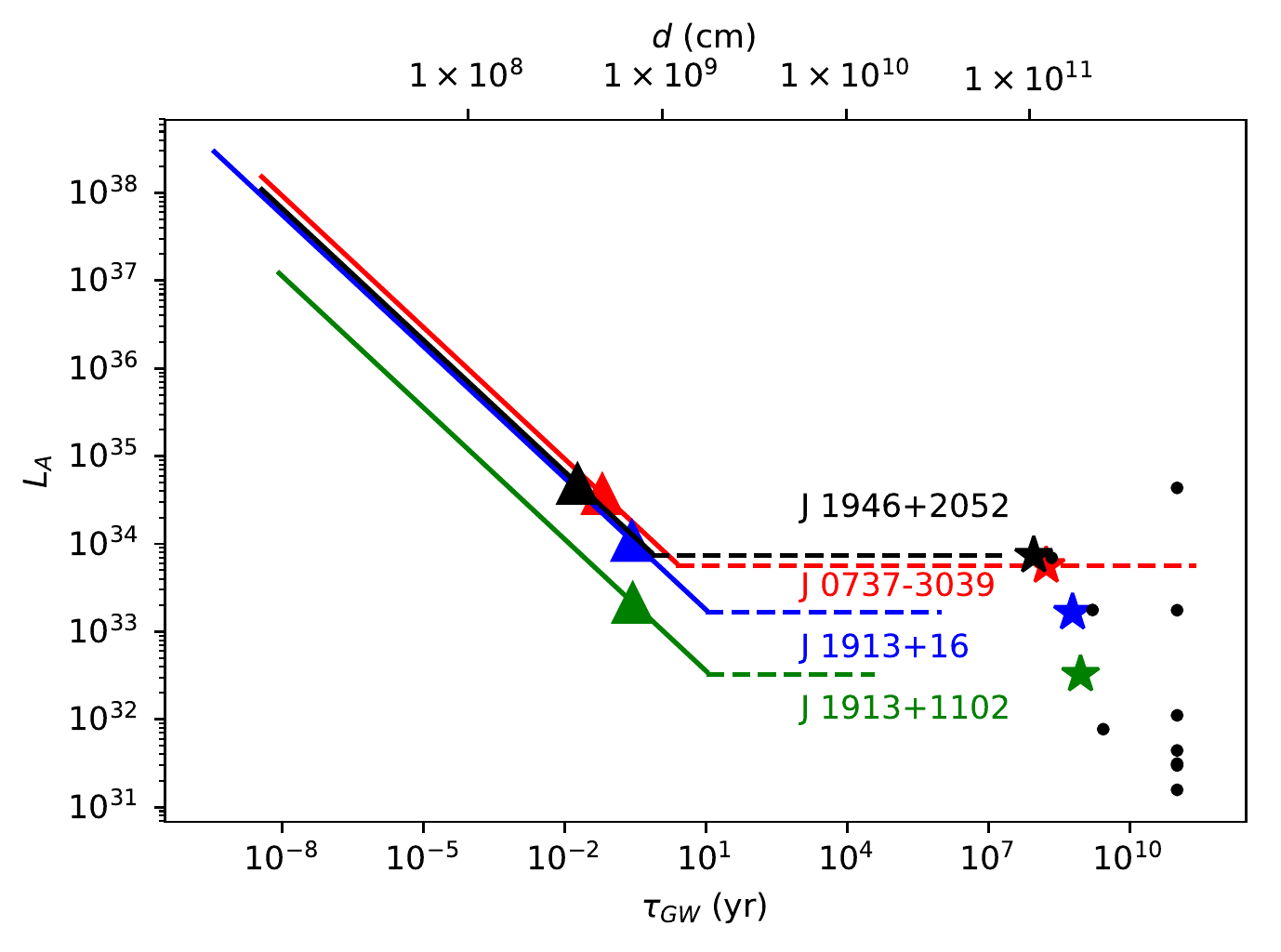}
\caption{The same as Figure \protect\ref{Figure:6}, where pulsar B has a magnetic field of $10^{12}$~G and a period of $0.3$~s and $1$~s, left and right panels respectively. }
\label{Figure:7}
\end{figure*}

As of now, there are 16 confirmed and candidate DNS systems, whose details are presented in Table \ref{tab:Pulsars}. The properties of both members are only known in the eponymous double pulsar J0737-3039A/B \citation{Stairs:2004} whose orbital period is $2.45$~hours and it has a gravitational wave decay time of 86 Myrs. J1946+2052 is an even more compact source with an orbital period of $1.88$~hours and a gravitational decay time of 46 Myr \citep{Stovall:2018}. Among the other 14 systems, in three of them, J1755-2550, J1906+0746 and J1807-2500B,  it is not yet confirmed whether their companions are neutron stars or white dwarfs \citep{Ng:2018, vanLeeuwen:2015, Yang:2017, Lynch:2012}. 

We have used the properties from four DNSs: J0737-3039A/B, J1946+2052, J1913+1102, B1913+16 to estimate how long before their merger they will exhibit any type of magnetospheric interaction and evaluate the potential observables. Except from J0737-3039A/B for which the properties of both pulsars are known, we had to make an assumption for the magnetic field strength and period of pulsar B for the other three systems. We set $P_{B} =3$s and we explore magnetic field strengths for pulsar B equal to $10^{12}$G, $5\times 10^{12}$G, $10^{13}$G and $5\times 10^{13}$G, shown on Figure \ref{Figure:6} top left, top right, bottom left and bottom right panels respectively. Then, we set the magnetic field of pulsar B to $10^{12}$G and the period $0.3$s and $1$s, Figure \ref{Figure:7} left and right panels repsectively. The actual properties of the double pulsar J0737-3039A/B are used in all plots in Figures \ref{Figure:6} and \ref{Figure:7}. The magnetic dipole strength quoted is derived using the vacuum dipole formula $B/{\rm G}=3.2\times 10^{19}\sqrt{(P/{\rm s})\dot{P}}$. The merger time due to gravitational wave radiation $\tau_{GW}$ are related to the separation of the two pulsar $d$ by 
\begin{eqnarray}
\tau_{GW}=\frac{5}{256}\frac{c^5}{G^{3}}\frac{d^{4}}{M_1M_2(M_1+M_2)}\,,
\end{eqnarray}
where $G$ is Newton's gravitational constant, and $M_1$, $M_2$ the masses of the members of the binary. 

\subsection{The double pulsar J0733-3039A/B}

Let us first consider the double pulsar J0737-3039A/B. This system is currently in the wind-magnetosphere interaction stage where the wind of pulsar A modifies the magnetosphere of pulsar B  \citep{Gourgouliatos:2011b, Perera:2012, Lomiashvili:2014}. Indeed, while the main source of X-ray radiation is pulsar A \citep{Chatterjee:2007}, there is an observable modulation due to pulsar B \citep{Pellizzoni:2008}. Subsequent analysis has confirmed not only that part of the  X-ray emission has a periodicity equal to that of pulsar B,  but also that it is orbitally modulated \citep{Iacolina:2016}. The X-ray emission from pulsar B is estimated to be $>2\times 10^{30}$erg~s$^{-1}$, which is 1.25 times higher than the spin-down power of pulsar B. Thus, it is impossible for pulsar B to power this radiation via its own spin-down \citep{Egron:2017}. On the contrary the spin-down power of A is  $6\times10^{33}$erg~s$^{-1}$, therefore, approximately $10^{-3}$ of the spin-down power of A is modulated by pulsar B. Following the analysis of section \ref{Application}, the system will be in the wind-magnetosphere interaction phase during most of its life, until approximately 1 yr before merger when it will enter the coupled magnetosphere stage. Then, the spin-down power of pulsar A will increase by about 4 orders magnitude. During its final days the system may undergo a Dungey-sphere state releasing rapidly large amounts of energy due to changes in the topology of the magnetosphere. Finally, the unipolar inductor will become possible only during $0.5$ s prior coalescence. 

\subsection{Galactic DNSs}

With respect to the other systems for which the magnetic field of pulsar B is unknown, we find that the results are sensitive to the strength of the magnetic field of B. A magnetic field of $10^{12}$G leads to results similar to the double pulsar.  Stronger magnetic fields for pulsar B ($5\times 10^{13}$G) are responsible for two main differences. Firstly, the wind-magnetosphere interaction starts later or may not occur at all. This is because the magnetic energy density of pulsar B is strong enough not to be affected by the wind of A as much as in the case of a mildly magnetised neutron star. Secondly, there is an earlier onset of the coupled magnetosphere phase, starting  $\sim10^{3}$~yr before the merger. As in the weakly magnetised case, the increase of the spin-down power of A rises by up to 4 orders of magnitude. During the final 20 years before coalescence, the system will fulfil the necessary conditions for a Dungey-sphere solution. Regarding the unipolar inductor phase, thought, the system will be capable of activating this mechanism for only a couple of seconds before merger.  

A combination of a shorter period ($0.3$~s) and a $10^{12}$G magnetic field for pulsar B leads to configurations where the wind-magnetosphere phase is suppressed. This is because the light-cylinder of B is rather small, therefore the magnetic field of B there is much stronger compared to the wind of A, Figure \ref{Figure:7} left panel. A longer period ($1$~s) for pulsar B permits a wind-magnetosphere interaction phase whose duration depends strongly on the properties of pulsar A. 

Considering J1946+2052 we expect it to be in the wind-magnetosphere interaction phase if its companion is a relatively slowly rotating mildly magnetised pulsar. Should they interact, we could expect modulation of the X-ray emission of pulsar A in the period of B as in the double pulsar. Given that no X-ray source has been identified yet with J1946+2052, this question could be resolved using through deep X-ray observations.

The estimated merger rate for galactic DNSs is $R_{g}=21_{-14}^{+28}$ Myr$^{-1}$ \citep{Kim:2015}. The anticipated number of galactic DNSs in the phases of wind-magnetosphere interaction and linked magnetospheres is a sensitive function of the magnetic fields and spin periods of the pulsars. If the double pulsar J0737-3039A/B is a characteristic example of the population we anticipate that there will be $\sim 2\times 10^{-5}$ galactic DNSs with a linked magnetosphere configuration. Should DNSs with $5\times 10^{13}$~G magnetic field for pulsar B were common the wind-magnetosphere interaction would be less frequent. In this case, the typical time spent in the linked magnetosphere phase would be $10^{3}$, as permitted by such a combination of magnetic field and period. We would then anticipate $2\times 10^{-2}$ DNSs at this stage to exist in the galaxy. Thus, such sources would be likely to exist in the local universe. 

\subsection{Implications for FRBs}

While the Dungey-sphere and unipolar inductor phases do not last long, they release large amounts of energy.  A magnetosphere altering its topology between closed and open states, could release amounts of energy up to a fraction of the magnetospheric energy during these final stages before merger. For a millisecond pulsar the energy contained in the Dungey-sphere is $\sim 2\times 10^{37} (B_A/10^{10}{\rm G})^2$ erg, a value close to the energy range of FRBs. Given the large uncertainty in distance the actual power FRBs cannot be determined accurately. Nevertheless, taking into account the cosmological origin their energy ranges from $10^{38}$ to $10^{41}$ erg \citep{Lorimer:2007, Thornton:2013, Bannister:2017}. Thus, the release of energy by switching from an open magnetosphere to a Dugney-sphere may be marginally powerful for an FRB. More interestingly though, this switching is not a one-off event. The separation of a DNS can be such to allow this solution for a up to a few years if the magnetic field of pulsar B is $5\times10^{13}$ G, or a few days for a magnetic field of $10^{12}$G. Taking into account the DNS merger rate $R=1540^{+3200}_{-1220}$ Gpc$^{-3}$yr$^{-1}$ \citep{Abbott:2017a}, we expect the number density of DNSs in the Dungey-sphere phase to be within the $10-10^{3}$Gpc$^{-3}$. A DNS whose separation is $10^{9}$~cm has an orbital period of $10$ s, thus it is likely that orientations favouring the transition between open and closed magnetospheres could occur every few periods. We remark though, that this interaction is highly non-trivial and requires a 3-D study of the magnetospheres.

\subsection{Pulsar spin evolution}

In our discussion we have taken for granted that the spin properties of the neutron stars do not vary as their orbit shrinks. \cite{Bildsten:1992} have shown that the inspiral is so fast that tidal interactions cannot affect the rotation of the members of the binary. Here we show that the magnetic interaction cannot have a drastic effect on the system either. 

Confining the magnetosphere of pulsar A leads to an increase of its spin-down power up to 4 orders of magnitude compared to their vacuum spin-down power $L_{A}<10^{4}L_{A,VAC}$. Since $L_{A}\propto P_{A}^{-3}\dot{P}_{A}$ a similar increase is expected to the period derivative. As the characteristic ages of the recycled millisecond pulsars are all above $10^{8}$~yr, even in the most optimistic scenario they will be scaled down to $10^{5}$~yr. Since the pulsar is in the enhanced $L_{A}$ phase for up to$10^{3}$ yr, this effect is not expected to have any major implications for their spin evolution. 

\begin{comment}
Similarly the effect because of the angular momentum transferred from pulsar B to A is minimal. The loss of angular momentum from pulsar A is $\Delta H_{A}=2\pi I\dot{P}_{A}P_{A}^{-2}\Delta t=\eta I\dot{P}_{A,VAC}P_{A}^{-2}\Delta t$, were $I$ is the moment of inertia of a neutron star, $\dot{P}_{A,VAC}$ is the period derivative corresponding to vacuum dipole spin-down, $\Delta t$ is the time during which the enhanced spin-down lasts and $\eta$ is the multiplication factor on the spin-down power of pulsar A because of magnetic field confinement. The angular momentum of pulsar B is $H_{B}=2\pi I P_{B}^{-1}$. Assuming that a fraction $\epsilon\leq 1$ of $\Delta H_{A}$ is transferred to pulsar B during the linked magnetosphere phase, it will need $H_{B}\sim \epsilon \Delta H_{A}$ for it to have some impact, which by virtue of the above expression becomes: $\epsilon \eta \Delta t \sim 2 \pi P_{A}^{2} \dot{P}_{A,VAC}^{-1}P_{B}^{-1}$. Considering a millisecond pulsar with $P_{A}=10$ms, $\dot{P}_{A,VAC}=10^{-19}$ and $P_{B}=1$s we obtain: $P_{A}^{2} \dot{P}_{A,VAC}^{-1} P_{B}^{-1}=3\times 10^{7}$ yr. The enhanced spin-down phase can last up to $10^{3}$yr, therefore one would need $\epsilon \eta=10^{4}$ for the angular momentum from A to B to have some impact. However, such a high value of $\eta$ is possible for only a very short time, approximately a few days before merger. Therefore, the magnetic coupling cannot transfer enough angular momentum to have a significant impact to the spin evolution of the members of the binary.
\end{comment}																																																																																																														
 
\section{Conclusion}

We have explored the structure of an axially symmetric rotating relativistic magnetosphere interacting with an externally imposed uniform magnetic field. We have found via analytical and numerical calculations, that one of the key elements of relativistically rotating pulsar magnetospheres, the equatorial current sheet, is either truncated or even completely suppressed if the externally imposed magnetic field is sufficiently strong. An interesting feature of these structures, it that for appropriately counter-aligned magnetic fields the system may adopt different types of equilibria, either in the form of twisted magnetic fields separated by current sheets from the external magnetospheres, or a fully confined Dungey-sphere solution in which no currents flow. Stronger external magnetic fields increase the fraction of open magnetic flux along which electric current flows. 

The implications of this interaction for pulsar magnetospheres could be potentially important and observable when considering DNSs. Here the role of the background field is played by the young pulsar which is expected to have a stronger magnetic field and a longer period. The magnetosphere of the millisecond pulsar is modified by this field once they are close enough. Even in the optimistic scenario where the young pulsar is  strongly magnetised ($5\times 10^{13}$G) and spins slowly ($3$s) such systems would be rare:  the probability of finding such a system in the Galaxy would be $\sim 0.02$. Finally, a system interchanging states between closed and an open magnetosphere may release significant amounts of energy. Order of magnitude estimates place the available energy close to the low estimates for FRBs.

The results presented here are valid based on the assumption of axisymmetry and stationarity. Realistic systems will have a combination of orientations, with orbital, spin and magnetic moment axes misaligned, while these evolve dynamically in time. While, the conclusions of the axisymmetric system are not directly transferable to the three-dimensional case, they still provide useful insight for more realistic cases.

\section*{Acknowledgements}
KNG would like to thank Ruth Lynden-Bell for hospitality in Cambridge during the summer of 2014 where the analytical and first numerical results were derived. KNG is grateful to an anonymous referee for their constructive comments on the applications of the analytical and numerical solutions to double neutron star systems.

\bibliographystyle{mnras}
\bibliography{Bibtex.bib}

\begin{thebibliography}{56}
\expandafter\ifx\csname natexlab\endcsname\relax\def\natexlab#1{#1}\fi

\bibitem[{Abbott} et~al.(2017{\natexlab{a}}){Abbott}, {Abbott}, {Abbott}
  et~al.]{Abbott:2017a}
{Abbott} B.~P., {Abbott} R., {Abbott} T.~D., et~al., 2017{\natexlab{a}},
  Physical Review Letters, 119, 16, 161101

\bibitem[{Abbott} et~al.(2017{\natexlab{b}}){Abbott}, {Abbott}, {Abbott}
  et~al.]{Abbott:2017b}
{Abbott} B.~P., {Abbott} R., {Abbott} T.~D., et~al., 2017{\natexlab{b}}, \apjl,
  850, L40

\bibitem[{Alexander} et~al.(2017){Alexander}, {Berger}, {Fong}
  et~al.]{Alexander:2017}
{Alexander} K.~D., {Berger} E., {Fong} W., et~al., 2017, \apjl, 848, L21

\bibitem[{Bannister} et~al.(2017){Bannister}, {Shannon}, {Macquart}
  et~al.]{Bannister:2017}
{Bannister} K.~W., {Shannon} R.~M., {Macquart} J.-P., et~al., 2017, \apjl, 841,
  L12

\bibitem[{Bhattacharya} \& {van den Heuvel}(1991)]{Bhattacharya:1991}
{Bhattacharya} D., {van den Heuvel} E.~P.~J., 1991, \physrep, 203, 1

\bibitem[{Bildsten} \& {Cutler}(1992)]{Bildsten:1992}
{Bildsten} L., {Cutler} C., 1992, \apj, 400, 175

\bibitem[{Chatterjee} et~al.(2007){Chatterjee}, {Gaensler}, {Melatos},
  {Brisken} \& {Stappers}]{Chatterjee:2007}
{Chatterjee} S., {Gaensler} B.~M., {Melatos} A., {Brisken} W.~F., {Stappers}
  B.~W., 2007, \apj, 670, 1301

\bibitem[{Contopoulos}(2005)]{Contopoulos:2005}
{Contopoulos} I., 2005, \aap, 442, 579

\bibitem[{Contopoulos} et~al.(1999){Contopoulos}, {Kazanas} \&
  {Fendt}]{Contopoulos:1999}
{Contopoulos} I., {Kazanas} D., {Fendt} C., 1999, \apj, 511, 351

\bibitem[{Cowperthwaite} et~al.(2017){Cowperthwaite}, {Berger}, {Villar}
  et~al.]{Cowperthwaite:2017}
{Cowperthwaite} P.~S., {Berger} E., {Villar} V.~A., et~al., 2017, \apjl, 848,
  L17

\bibitem[{Dungey}(1961)]{Dungey:1961}
{Dungey} J.~W., 1961, Physical Review Letters, 6, 47

\bibitem[{Egron} et~al.(2017){Egron}, {Pellizzoni}, {Pollock}
  et~al.]{Egron:2017}
{Egron} E., {Pellizzoni} A., {Pollock} A., et~al., 2017, \apj, 838, 120

\bibitem[{Etienne} et~al.(2012){Etienne}, {Paschalidis} \&
  {Shapiro}]{Etienne:2012}
{Etienne} Z.~B., {Paschalidis} V., {Shapiro} S.~L., 2012, \prd, 86, 8, 084026

\bibitem[{Goldreich} \& {Julian}(1969)]{Goldreich:1969a}
{Goldreich} P., {Julian} W.~H., 1969, \apj, 157, 869

\bibitem[{Goldreich} \& {Lynden-Bell}(1969)]{Goldreich:1969b}
{Goldreich} P., {Lynden-Bell} D., 1969, \apj, 156, 59

\bibitem[{Gourgouliatos} \& {Lynden-Bell}(2011)]{Gourgouliatos:2011a}
{Gourgouliatos} K.~N., {Lynden-Bell} D., 2011, \mnras, 410, 257

\bibitem[{Gourgouliatos} et~al.(2011){Gourgouliatos}, {Lyutikov}, {Lomiashvili}
  \& {Perera}]{Gourgouliatos:2011b}
{Gourgouliatos} K.~N., {Lyutikov} M., {Lomiashvili} D., {Perera} B.~B.~P.,
  2011, in { American Institute of Physics Conference Series\/}, edited by
  M.~{Burgay}, N.~{D'Amico}, P.~{Esposito}, A.~{Pellizzoni}, A.~{Possenti},
  vol. 1357 of { American Institute of Physics Conference Series\/},  304--305

\bibitem[{Gruzinov}(2005)]{Gruzinov:2005}
{Gruzinov} A., 2005, Physical Review Letters, 94, 2, 021101

\bibitem[{Hansen} \& {Lyutikov}(2001)]{Hansen:2001}
{Hansen} B.~M.~S., {Lyutikov} M., 2001, \mnras, 322, 695

\bibitem[{Iacolina} et~al.(2016){Iacolina}, {Pellizzoni}, {Egron}
  et~al.]{Iacolina:2016}
{Iacolina} M.~N., {Pellizzoni} A., {Egron} E., et~al., 2016, \apj, 824, 87

\bibitem[{Kalapotharakos} \& {Contopoulos}(2009)]{Kalapotharakos:2009}
{Kalapotharakos} C., {Contopoulos} I., 2009, \aap, 496, 495

\bibitem[{Kalapotharakos} et~al.(2012){Kalapotharakos}, {Contopoulos} \&
  {Kazanas}]{Kalapotharakos:2012}
{Kalapotharakos} C., {Contopoulos} I., {Kazanas} D., 2012, \mnras, 420, 2793

\bibitem[{Kim} et~al.(2015){Kim}, {Perera} \& {McLaughlin}]{Kim:2015}
{Kim} C., {Perera} B.~B.~P., {McLaughlin} M.~A., 2015, \mnras, 448, 928

\bibitem[{Komissarov}(2006)]{Komissarov:2006}
{Komissarov} S.~S., 2006, \mnras, 367, 19

\bibitem[{Kruskal} et~al.(1958){Kruskal}, {Johnson}, {Gottlieb} \&
  {Goldman}]{Kruskal:1958}
{Kruskal} M.~D., {Johnson} J.~L., {Gottlieb} M.~B., {Goldman} L.~M., 1958,
  Physics of Fluids, 1, 421

\bibitem[{Lomiashvili} \& {Lyutikov}(2014)]{Lomiashvili:2014}
{Lomiashvili} D., {Lyutikov} M., 2014, \mnras, 441, 690

\bibitem[{Lorimer} et~al.(2007){Lorimer}, {Bailes}, {McLaughlin}, {Narkevic} \&
  {Crawford}]{Lorimer:2007}
{Lorimer} D.~R., {Bailes} M., {McLaughlin} M.~A., {Narkevic} D.~J., {Crawford}
  F., 2007, Science, 318, 777

\bibitem[{Lynch} et~al.(2012){Lynch}, {Freire}, {Ransom} \&
  {Jacoby}]{Lynch:2012}
{Lynch} R.~S., {Freire} P.~C.~C., {Ransom} S.~M., {Jacoby} B.~A., 2012, \apj,
  745, 109

\bibitem[{Lynden-Bell}(2003)]{Lynden-Bell:2003}
{Lynden-Bell} D., 2003, \mnras, 341, 1360

\bibitem[{Margutti} et~al.(2017){Margutti}, {Berger}, {Fong}
  et~al.]{Margutti:2017}
{Margutti} R., {Berger} E., {Fong} W., et~al., 2017, \apjl, 848, L20

\bibitem[{Metzger} \& {Berger}(2012)]{Metzger:2012}
{Metzger} B.~D., {Berger} E., 2012, \apj, 746, 48

\bibitem[{Metzger} et~al.(2010){Metzger}, {Mart{\'{\i}}nez-Pinedo}, {Darbha}
  et~al.]{Metzger:2010}
{Metzger} B.~D., {Mart{\'{\i}}nez-Pinedo} G., {Darbha} S., et~al., 2010,
  \mnras, 406, 2650

\bibitem[{Metzger} \& {Zivancev}(2016)]{Metzger:2016}
{Metzger} B.~D., {Zivancev} C., 2016, \mnras, 461, 4435

\bibitem[{Ng} et~al.(2018){Ng}, {Kruckow}, {Tauris} et~al.]{Ng:2018}
{Ng} C., {Kruckow} M.~U., {Tauris} T.~M., et~al., 2018, \mnras, 476, 4315

\bibitem[{Nicholl} et~al.(2017){Nicholl}, {Berger}, {Kasen}
  et~al.]{Nicholl:2017}
{Nicholl} M., {Berger} E., {Kasen} D., et~al., 2017, \apjl, 848, L18

\bibitem[{Palenzuela} et~al.(2013){Palenzuela}, {Lehner}, {Ponce}
  et~al.]{Palenzuela:2013}
{Palenzuela} C., {Lehner} L., {Ponce} M., et~al., 2013, Physical Review
  Letters, 111, 6, 061105

\bibitem[{Paschalidis} et~al.(2013){Paschalidis}, {Etienne} \&
  {Shapiro}]{Paschalidis:2013}
{Paschalidis} V., {Etienne} Z.~B., {Shapiro} S.~L., 2013, \prd, 88, 2, 021504

\bibitem[{Pellizzoni} et~al.(2008){Pellizzoni}, {Tiengo}, {De Luca}, {Esposito}
  \& {Mereghetti}]{Pellizzoni:2008}
{Pellizzoni} A., {Tiengo} A., {De Luca} A., {Esposito} P., {Mereghetti} S.,
  2008, \apj, 679, 664

\bibitem[{Perera} et~al.(2012){Perera}, {Lomiashvili}, {Gourgouliatos},
  {McLaughlin} \& {Lyutikov}]{Perera:2012}
{Perera} B.~B.~P., {Lomiashvili} D., {Gourgouliatos} K.~N., {McLaughlin} M.~A.,
  {Lyutikov} M., 2012, \apj, 750, 130

\bibitem[{Piro}(2012)]{Piro:2012}
{Piro} A.~L., 2012, \apj, 755, 80

\bibitem[{Ponce} et~al.(2014){Ponce}, {Palenzuela}, {Lehner} \&
  {Liebling}]{Ponce:2014}
{Ponce} M., {Palenzuela} C., {Lehner} L., {Liebling} S.~L., 2014, \prd, 90, 4,
  044007

\bibitem[{Press} et~al.(1992){Press}, {Teukolsky}, {Vetterling} \&
  {Flannery}]{Press:1992}
{Press} W.~H., {Teukolsky} S.~A., {Vetterling} W.~T., {Flannery} B.~P., 1992,
  {Numerical recipes in FORTRAN. The art of scientific computing}

\bibitem[{Rezzolla} et~al.(2011){Rezzolla}, {Giacomazzo}, {Baiotti}, {Granot},
  {Kouveliotou} \& {Aloy}]{Rezzolla:2011}
{Rezzolla} L., {Giacomazzo} B., {Baiotti} L., {Granot} J., {Kouveliotou} C.,
  {Aloy} M.~A., 2011, \apjl, 732, L6

\bibitem[{Shafranov}(1958)]{Shafranov:1958}
{Shafranov} V.~D., 1958, Soviet Journal of Experimental and Theoretical
  Physics, 6, 545

\bibitem[{Spitkovsky}(2006)]{Spitkovsky:2006}
{Spitkovsky} A., 2006, \apjl, 648, L51

\bibitem[{Stairs}(2004)]{Stairs:2004}
{Stairs} I.~H., 2004, Science, 304, 547

\bibitem[{Stovall} et~al.(2018){Stovall}, {Freire}, {Chatterjee}
  et~al.]{Stovall:2018}
{Stovall} K., {Freire} P.~C.~C., {Chatterjee} S., et~al., 2018, \apjl, 854, L22

\bibitem[{Tauris} et~al.(2017){Tauris}, {Kramer}, {Freire} et~al.]{Tauris:2017}
{Tauris} T.~M., {Kramer} M., {Freire} P.~C.~C., et~al., 2017, \apj, 846, 170

\bibitem[{Thornton} et~al.(2013){Thornton}, {Stappers}, {Bailes}
  et~al.]{Thornton:2013}
{Thornton} D., {Stappers} B., {Bailes} M., et~al., 2013, Science, 341, 53

\bibitem[{Timokhin}(2006)]{Timokhin:2006}
{Timokhin} A.~N., 2006, \mnras, 368, 1055

\bibitem[{Troja} et~al.(2010){Troja}, {Rosswog} \& {Gehrels}]{Troja:2010}
{Troja} E., {Rosswog} S., {Gehrels} N., 2010, \apj, 723, 1711

\bibitem[{Tsang}(2013)]{Tsang:2013}
{Tsang} D., 2013, \apj, 777, 103

\bibitem[{Tsang} et~al.(2012){Tsang}, {Read}, {Hinderer}, {Piro} \&
  {Bondarescu}]{Tsang:2012}
{Tsang} D., {Read} J.~S., {Hinderer} T., {Piro} A.~L., {Bondarescu} R., 2012,
  Physical Review Letters, 108, 1, 011102

\bibitem[{van Leeuwen} et~al.(2015){van Leeuwen}, {Kasian}, {Stairs}
  et~al.]{vanLeeuwen:2015}
{van Leeuwen} J., {Kasian} L., {Stairs} I.~H., et~al., 2015, \apj, 798, 118

\bibitem[{Vlahakis}(2004)]{Vlahakis:2004}
{Vlahakis} N., 2004, \apj, 600, 324

\bibitem[{Yang} et~al.(2017){Yang}, {Zhang}, {Li} et~al.]{Yang:2017}
{Yang} Y.-Y., {Zhang} C.-M., {Li} D., et~al., 2017, \apj, 835, 185

\end{thebibliography}

\label{lastpage}

\end{document}